\documentclass[10pt, letterpaper, conference]{IEEEtran}

\pdfoutput=1

\def\remarkscolor{orange}

\def\isanonymous{0} 

\def\isextended{1}
\def\istvt{0}  

\if \isextended 0
\def\figurewidth{0.47}
\def\figurewidthnarrow{0.4}
\else
\def\figurewidth{0.75}
\def\figurewidthnarrow{0.45}
\fi

\ifCLASSINFOpdf
	\usepackage[pdftex]{graphicx}
	\graphicspath{{./figures/}}
\else
  \usepackage[dvips]{graphicx}
  \graphicspath{{./figures/}}
\fi

\usepackage[cmex10]{amsmath}
\usepackage[caption = false, font = footnotesize]{subfig}
\usepackage{amsthm}
\usepackage{amsfonts}

\newtheorem{theorem}{Theorem}

\newcommand*{\Rom}[1]{\uppercase\expandafter{\romannumeral #1\relax}} 
\usepackage{textcomp}
\usepackage{gensymb} 
\usepackage{bbm}
\usepackage{authblk}
\usepackage{stackengine}
\usepackage{changes} 
\setaddedmarkup{{\color{\remarkscolor} #1}} 
\usepackage{multirow} 

\DeclareMathOperator*{\unif}{unif}
\DeclareMathOperator*{\E}{\mathrm{E}}

\newcommand{\overbar}[1]{\mkern 1.5mu\overline{\mkern-1.5mu#1\mkern-1.5mu}\mkern 1.5mu}
\DeclareMathOperator*{\Real}{\mathbb{R}}
\DeclareMathOperator*{\Natural}{\mathbb{N}}
\DeclareMathOperator*{\indicator}{\mathbbm{1}}
\DeclareMathOperator*{\Realsquare}{\mathbb{R}^2}

\DeclareMathOperator*{\rsu}{rsu}
\DeclareMathOperator*{\vehicle}{v}

\DeclareMathOperator*{\interest}{interest} 
\DeclareMathOperator*{\communication}{comm} 
\DeclareMathOperator*{\gap}{gap} 
\DeclareMathOperator*{\Prob}{P} 
\DeclareMathOperator*{\uplink}{UL} 
\DeclareMathOperator*{\downlink}{DL} 
\DeclareMathOperator*{\vtoi}{V2I} 
\DeclareMathOperator*{\vtov}{V2V} 

\DeclareMathOperator*{\broadcast}{B}
\DeclareMathOperator*{\unicast}{U}
\DeclareMathOperator*{\front}{front} 
\DeclareMathOperator*{\meter}{m}

\DeclareMathOperator*{\road}{road}
\DeclareMathOperator*{\void}{void} 
\DeclareMathOperator*{\HPPP}{HPPP} 
\DeclareMathOperator*{\object}{o} 

\def\delequal{\mathrel{\ensurestackMath{\stackon[1pt]{=}{\scriptstyle\Delta}}}} 

\theoremstyle{definition}
\newtheorem{definition}{Definition}
\theoremstyle{statement}

\title{Performance and Scaling of Collaborative Sensing 
	and Networking for Automated Driving 
	Applications
	}
\if\isanonymous1\author{}\else
\author[*]{Yicong Wang}
\author[*]{Gustavo de Veciana}
\author[$\dagger$]{Takayuki Shimizu}
\author[$\dagger$]{Hongsheng Lu}
\affil[*]{Department of Electrical and Computer Engineering, The University of Texas at Austin}
\affil[$\dagger$]{TOYOTA InfoTechnology Center, U.S.A., Inc., Mountain View, CA}
\fi
\begin{document}
\maketitle
\thispagestyle{plain}
\pagestyle{plain}

\begin{abstract}
        A critical requirement for automated driving systems is enabling
        situational awareness in dynamically changing environments.
        To that end vehicles will be equipped with diverse sensors,
        e.g., LIDAR, cameras, mmWave radar, etc.
        Unfortunately the sensing `coverage' is limited by environmental obstructions,
        e.g., other vehicles, buildings, people,
        objects etc.
        A possible solution is to adopt collaborative sensing
        amongst vehicles possibly assisted by infrastructure.
        This paper introduces new models and performance analysis for vehicular
        collaborative sensing and networking. In particular, coverage gains are quantified, as are
        their dependence on the penetration of vehicles participating in collaborative sensing. 
        We also evaluate the associated communication loads 
        in terms of the Vehicle-to-Vehicle (V2V) and Vehicle-to-Infrastructure (V2I)
        capacity requirements and
        how these depend on penetration.
        We further explore how collaboration with sensing capable 
        infrastructure improves sensing performance, as well as the benefits in
        utilizing spatio-temporal dynamics,
        e.g., collaborating with vehicles moving in the opposite direction.
        Collaborative sensing is shown to greatly improve sensing performance, 
        e.g., improves coverage from $20\%$ to $80\%$ with a $20\%$ penetration. 
        In scenarios with limited penetration and high coverage requirements, 
        infrastructure can be used to both help sense the environment and relay data. 
        Once penetration is high enough, sensing vehicles provide good coverage 
        and data traffic can be effectively `offloaded' to V2V connectivity, 
        making V2I resources available to support other in-car services.
\end{abstract}

\IEEEpeerreviewmaketitle

\section{Introduction}
\label{section:introduction}

In future automated driving systems, vehicles will need to
maintain real-time situational awareness in dynamically changing
environments. 
Despite vehicles being equipped with multiple sensors, 
e.g., radar, LIDAR, cameras etc., the sensing `coverage'
of a single vehicle is limited. 
Indeed such sensors typically rely on a Line-Of-Sight (LOS) to 
detect and track objects, so their performance is fragile in 
obstructed environments, e.g., a vehicle may have limited visibility 
of what is happening several cars ahead of it. Such information 
could be needed for path planning, determining car-following 
distance, taking critical safety manouvers, etc.
Further without access to diverse points of view of 
an object, it may be difficult to quickly detect and recognize what it is, e.g., 
a cyclist viewed only from the front may look like a pedestrian.
 
To overcome this problem researchers and industry are considering 
enabling {\em distributed collaborative sensing} amongst neighboring 
vehicles, and possibly infrastructure, e.g., Road Side Units (RSUs) 
and/or base stations.
The idea is to enable automated vehicles to 
exchange High Resolution (HD) and/or processed data from vehicles 
and/or RSUs to enhance timely perception of the environment,
see e.g., \cite{kim2015multivehicle}\cite{hobert2015enhancements}.
The benefits of this approach will depend on the penetration
of collaborating vehicles/RSUs as well as the density and character of
obstructions in the environment.
The communication loads to share sensed information can be high and will need
to be met by enabling new forms of connectivity.

Collaborative sensing is likely to be one of key functionalities for cooperative
automated driving
\cite{hobert2015enhancements}, and one of the three most important use cases 
of future 5G systems \cite{5gppp2015automotive}. 
Thus a basic understanding of sensing performance and traffic scaling is 
of great interest. It may involve substantial data rates per vehicle,
e.g., $53$ Mbps, for highly automated driving, 
and require low end-to-end delays, 
e.g., $100$~ms or less depending on the use case \cite{3gpp2017v2xstage1}.
At high vehicle densities realizing such data exchanges via 
Vehicle-to-Infrastructure (V2I) resources 
is not likely to be possible, e.g., there could be tens to 
hundreds of vehicles sharing a base station. 
A possible solution is to leverage direct data exchanges amongst
vehicles. In particular short range millimeter wave (mmWave) based 
LOS Vehicle-to-Vehicle (V2V) links can support
exceedingly high data rates. Unfortunately such links are also susceptible to
obstructions, and thus, not unlike collaborative sensing itself, 
the connectivity of such V2V networks 
is limited by the penetration of vehicles with such communication
capabilities and obstructions in the environment.  
Thus in order to be viable (and reliable) collaborative sensing
applications will leverage a mix of V2V and V2I connectivity, likely
attempting to offload as much traffic as possible to the V2V networks.

The aim of this paper is to develop initial models and analysis of the 
benefits, communication loads and requirements for vehicular collaborative sensing and networking.
We focus on two intertwined classes of questions: \\
\emph{1. What are simple and tractable metrics for collaborative 
	sensing performance in obstructed environments? 
     How does performance scale with the penetration of collaborating 
     vehicles and density of obstructions?} \\
\emph{2. What are the network connectivity-capacity requirements to 
         support collaborative sensing on V2V/V2I networks as a function 
         of the penetration and density of vehicles?}\\
Note that while our focus will be on vehicular networks, other 
distributed autonomous systems built on wireless systems 
share similar characteristics, including, e.g., robotic or possibly emerging 
aerial drone applications. 

\emph{Contributions.}
The key contributions of this paper are as follows. 
\begin{itemize}
	\item We introduce a stochastic geometric model to study 
		collaborative sensing in obstructed environments along 
		with associated performance metrics capturing sensing coverage.
	\item We quantify the performance of collaborative sensing 
		for varying coverage requirements, 
		vehicle/object densities and penetration of collaborative sensing vehicles.
	\item We explore heterogeneous architectures of sensing and communication 
		combining vehicles and infrastructure.
		Our study on the sensing performance and capacity requirements 
		exhibits the critical role that infrastructure assistance might need to
		play in improving sensing coverage and providing reliable communication
		especially at the early stages of collaborative sensing at low penetrations.
	\item We show that exploiting spatio-temporal dynamics collaboration with
	flows of vehicles moving in different directions improves the performance of
	collaborative sensing, yet the benefit is limited at high penetrations
	of collaborating vehicles.
\end{itemize}
Our analytical results are based on simple/tractable models that capture the
essence of such systems. We further conduct simulations of typical
road traffic scenarios to validate our analysis and provide additional 
quantitative assessments.

\emph{Related work.}
Collaborative sensing is likely to be one of the key 
enabling technologies for automated driving systems. 
Vehicles can exchange real time sensor information with vehicles/RSUs to enhance
their view of an obstructed environment 
\cite{rauch2011analysis}\cite{li2011multi}\cite{rauch2012car2x}\cite{kim2013cooperative}\cite{zhao2017cooperative}.
An analysis of the scaling and performance of such systems has however not been done before. 

Currently available communication protocols such as 
Dedicated Short-Range Communication (DSRC) \cite{kenney2011dedicated}
have limited data rates, e.g., IEEE 802.11p supports 3--27 Mbps (typically 6Mbps).
LTE systems are evolving to support 
safety-related V2X applications \cite{seo2016lte}, but still 
provide limited capacity and face challenges
associated with the high densities of UEs.
To serve the requirements of collaborative sensing, 
3GPP defined various use cases and requirements in \cite{3gpp2017v2xstage1}\cite{3gpp5gv2x}.
Also mmWave technology is being considered to support 
the sharing of HD sensor data \cite{va2016millimeter}\cite{choi2016millimeter}.

The capacity of Vehicular Ad Hoc Networks (VANETs) has been studied in a variety of works, see e.g., 
\cite{zhang2012multicast}\cite{kwon2016analysis}\cite{he2017transmission}\cite{giang2016spatial}.
The communication requirements for collaborative sensing, i.e., each vehicle
requiring local many-to-many information sharing, 
is different from that typically considered in VANET studies where the source
and destination of data need not be close by. 
The existing capacity analysis needs to be adapted to this many-to-many setting.
The authors of \cite{ozbilgin2016evaluating} study the 
communication loads on a single vehicle, 
but obstructions and networking are not considered. 

To our knowledge, our work on modeling and assessing the performance of 
collaborative sensing is novel. It can be viewed as a stochastic version 
of what are referred to as the art gallery problem(s) \cite{o1987art}. 
These problems typically address questions such as the number and placement
of cameras/guards in a fixed environment to meet a pre-specified coverage criterion.
Hence this paper also contributes new results of this type but for random
sensor placement and obstructed environments. Such results are more appropriate
towards understanding vehicular systems ``in the wild''.

\emph{Organization.}
We begin by proposing a 2D model for sensing in obstructed 
environments in Section~\ref{section:model}. We then quantify 
the benefits that collaborative sensing would afford in terms of sensing 
coverage in Section~\ref{section:collaborative}. 
In Section~\ref{section:scalability} we analyze the capacity requirements 
on V2V and V2I networks.
In Section~\ref{section:dynamics} we study the performance of collaborative sensing
in the presence of spatio-temporal dynamics. 
We conclude the paper in Section~\ref{section:conclusion}.

\section{Modeling Collaborative Sensing in Obstructed Environments} \label{section:model}

We begin by introducing a simple stochastic geometric model to study the character of collaborative sensing.
\footnote{Note that we will focus on describing a 2D model although it can be
extended to 2.5D or 3D.}

\subsection{Obstructed Environments and Sensing Capabilities}\label{subsection:sensing:model}

The environment includes all objects, i.e., vehicles, 
pedestrians, buildings, etc. In some settings there may be 
substantial a priori knowledge regarding the environment, e.g., static elements 
that are part of a previously computed HD maps \cite{seif2016autonomous}. 
While the presence of such objects is already known they still 
impact collaborative sensing as they can obstruct a sensor's field of view, 
e.g., a building may obstruct a vehicle's view when entering an intersection.
For simplicity we shall not differentiate among static and dynamic objects, 
and focus on sensing at a snapshot in time\footnote{In practice collaborative sensing system will track objects over time. Thus taking the snapshot point of view can be considered ``worst case'' assumption.}.

\begin{figure}[hbt]
	\if\isextended0\vspace{-0.35in}\else\vspace{-0.3in}\fi
	\centering
	\subfloat[Model for object $i$.]{\label{subfig:model:environment:object}\includegraphics[width=0.4\columnwidth]{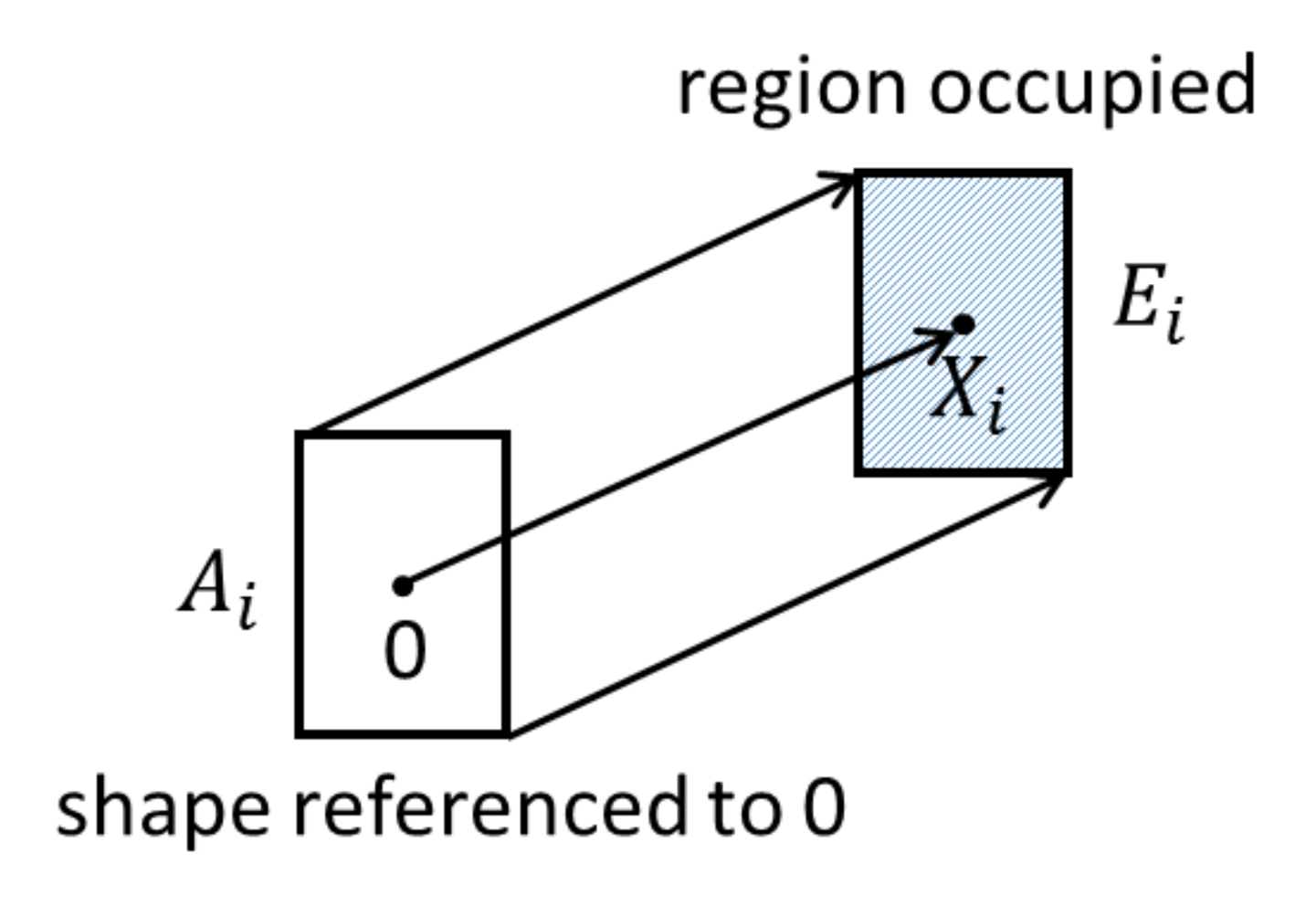}} 
	\hfill
	\subfloat[Model for environment.]{\label{subfig:model:environment:environment}
		\includegraphics[height=0.3\columnwidth]{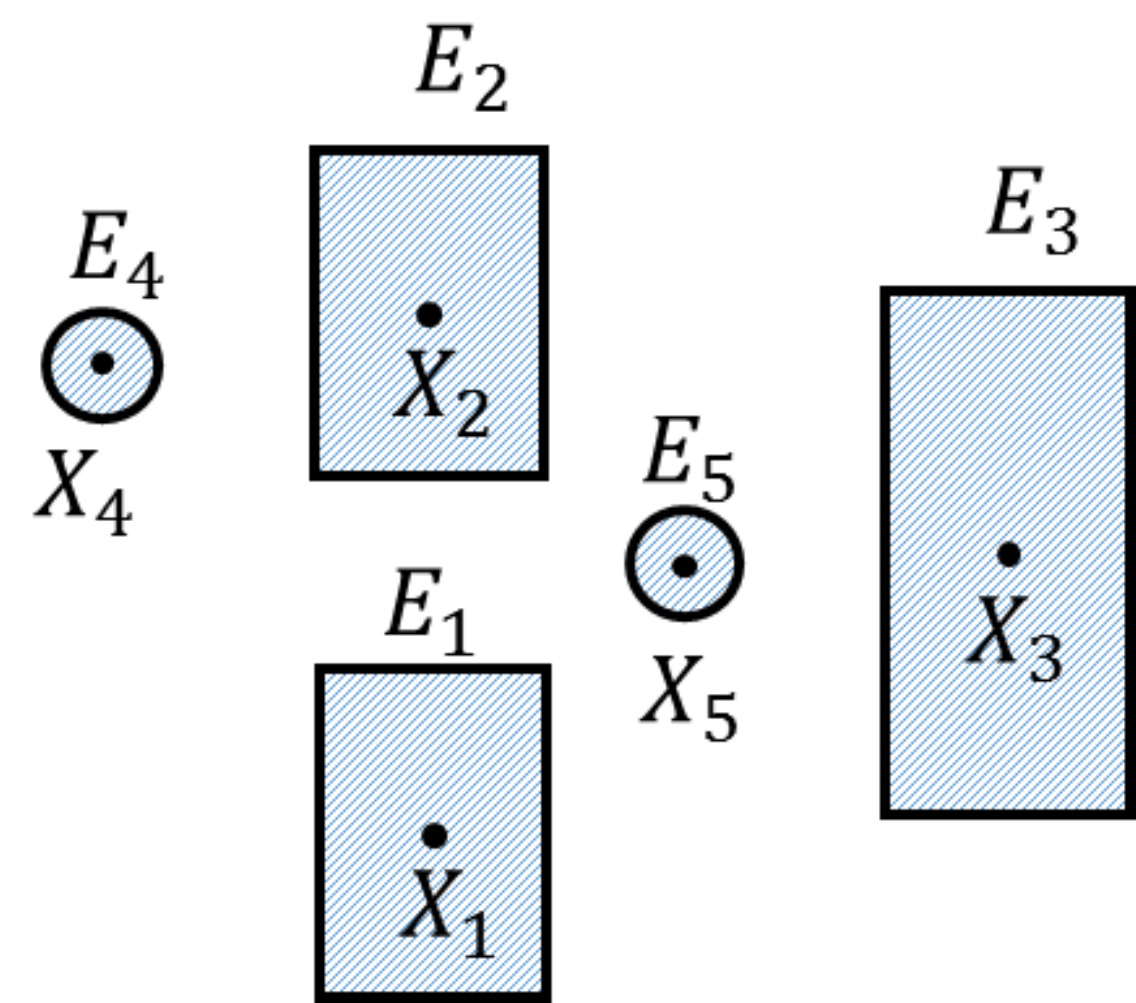}} 
	\hspace*{\fill}
	\caption[]{Model for environment based on randomly located and shaped objects.}
	\label{fig:model:environment}
\end{figure}

The centers of objects are located on 2-D plane
according to a Homogeneous Poisson Point Process (HPPP) $\Phi$ 
with intensity $\lambda$, i.e.,
$$
\Phi = \{X_i | X_i\in \Realsquare, ~i = {\Natural}^+\} \sim \mbox{HPPP}(\lambda),
$$
where $X_i$ is the location of object $i$, 
and $\Natural^+$ is the set of positive integers. 
Each object, say $i$, has a shape modeled by a random closed convex set denoted
$A_i\subset \Real^2$ referenced to the origin $0$ and independent of $X_i$. 
We let $E_i$ denote the region it occupies which is given by 
\begin{equation*}
E_i = \{X_i\}\oplus A_i \delequal \{X_i + x|x\in A_i\},
\end{equation*}
i.e., the object's shape $A_i$ shifted to its location $X_i$, 
where $\oplus$ is the Minkowski sum, see Fig.~\ref{subfig:model:environment:object}.
Thus $E = \overset{\infty}{\underset{i = 1}{\cup}}E_i$ denotes the region 
occupied by objects in the environment. 
We refer to the region not occupied by objects, $E^c = \{x|x\notin E\}$, as the \emph{void space}.
Fig.~\ref{subfig:model:environment:environment} illustrates our model for the 
environment. 

It is unavoidable that initially as automated driving technologies are 
introduced, only a fraction of vehicles will be equipped with sensors and/or 
participate in collaborative sensing.
Thus only the subset 
equipped with sensors can participate in collaborative sensing --
we shall refer to such objects as sensors. 
Each object has an independent probability $p_s$ of being a sensor. 
Thus the locations of sensors, $\Phi^s$, 
correspond to an independent thinning \cite{chiu2013stochastic} of $\Phi$, and
so $\Phi^s \sim $HPPP$(\lambda^s)$ where $\lambda_s = p_s\lambda$.
For such objects we assume for simplicity that each has \emph{one} sensor, and 
denote by $Y_i\in\Real^2$ the relative placement of the sensor 
on object $i$ referenced to $X_i$, so the location 
of sensor $i$ is given by $X_i + Y_i\in E_i$. 
Each sensor $i$ is assumed to have a radial sensing support $S_i^0\subset\Real^2$ 
referenced to the location of the sensor which is defined as follows.
\begin{definition}{(Radial sensing support)}
	The \emph{radial sensing support} of a sensor $i$ 
	referenced to the origin, $S_i^0$, is the set of locations
	that can be viewed if the sensor is located at $0$ and the LOS to the location is not obstructed.
	The set $S_i^0$ can be represented in polar coordinates as follows, 
	\begin{equation}
		S_i^0 = \big\{(r, \theta)\big| r\in [0, r_{\max}^i(\theta)] , \theta \in [0,2\pi] \big\},
	\end{equation}
	where $r_{\max}^i(\theta)$ is the maximum sensing range in direction $\theta$.
\end{definition}

\begin{figure}[hbt]
	\if\isextended0\vspace{-0.25in}\fi
	\centering
	\subfloat[Radial sensing support $S_i^0$]{\label{subfig:model:radial_support_0}\includegraphics[height=0.25\columnwidth]{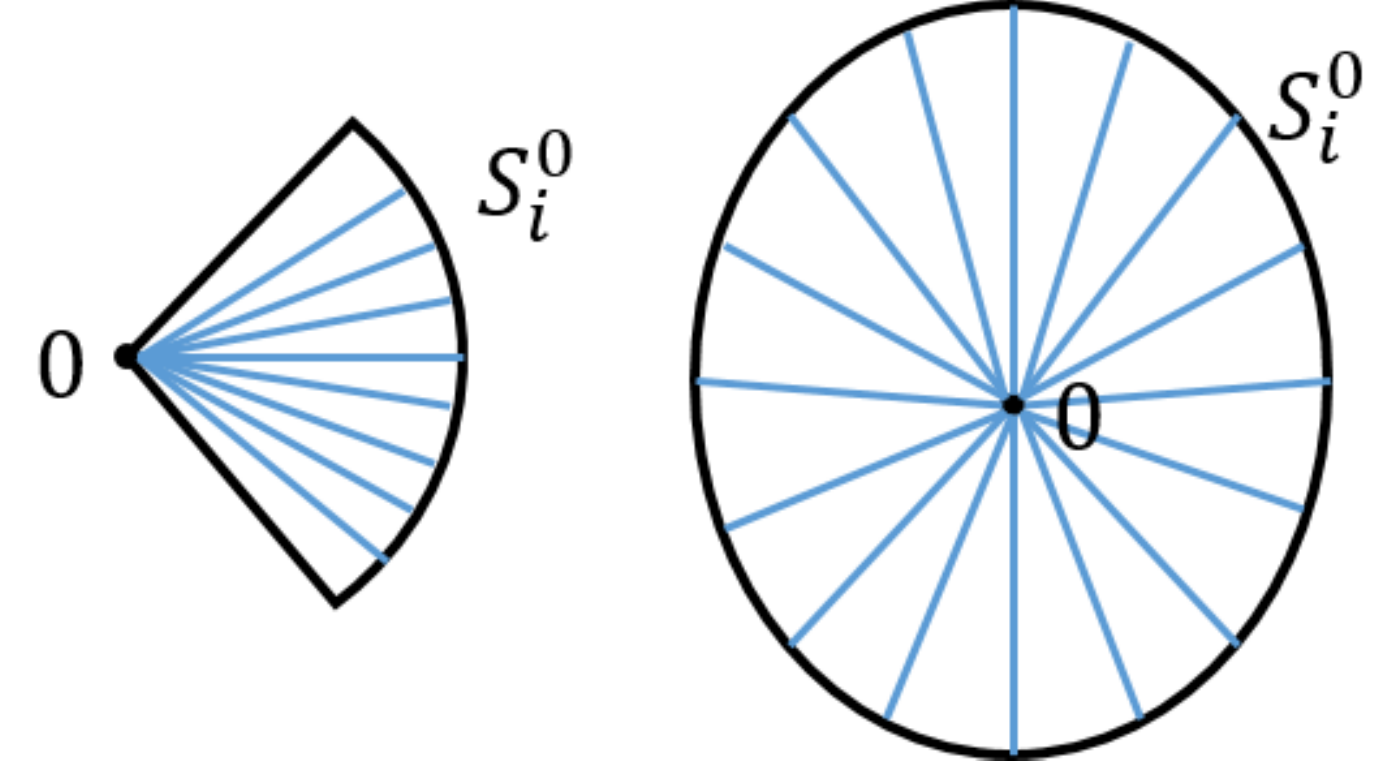}}
	\hfill
	\subfloat[Sensing support $S_i$]{\label{subfig:model:radial_support}
		\includegraphics[height=0.25\columnwidth]{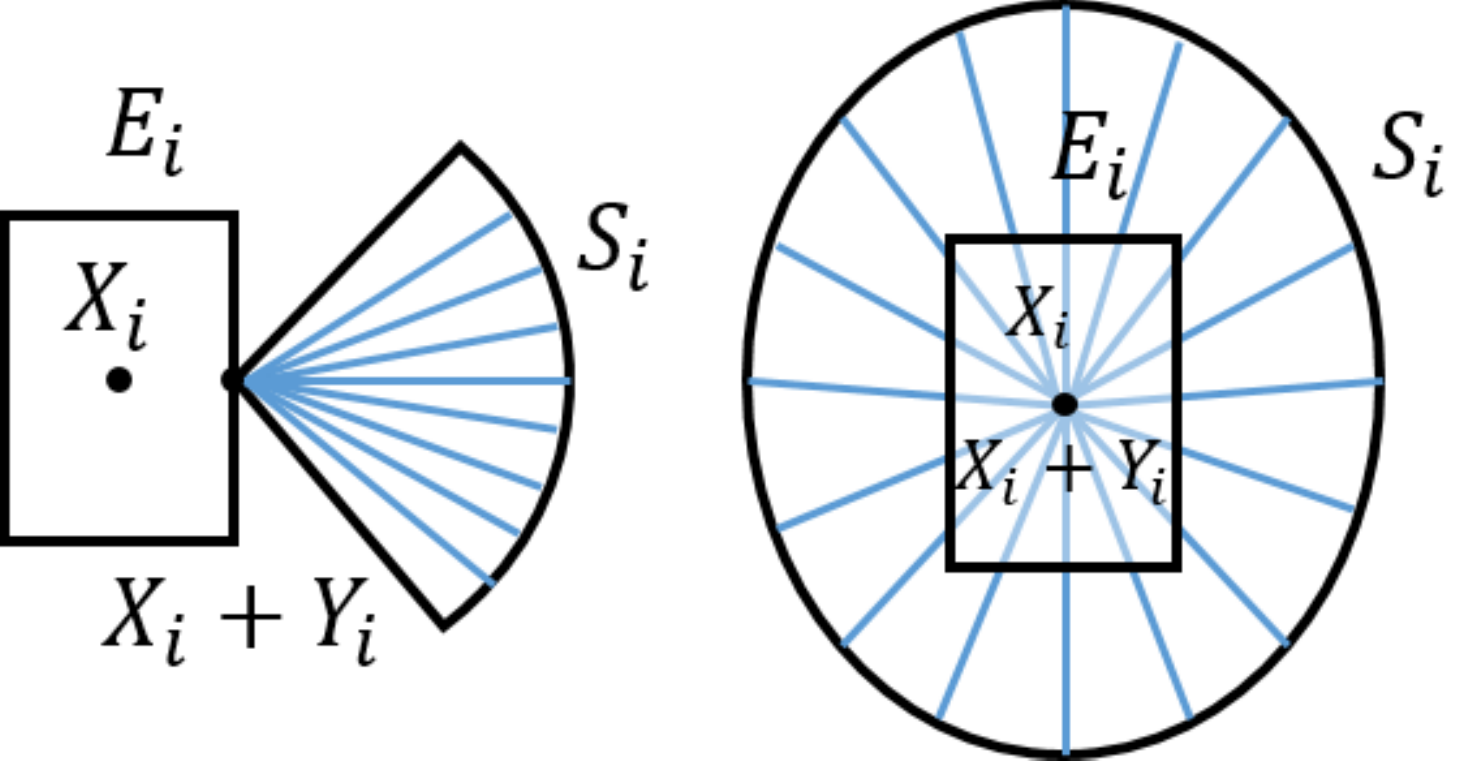}} \hfill
	\caption[]{\subref{subfig:model:environment:object} Radial support referenced to the 
			origin $0$ and \subref{subfig:model:radial_support} the sensing support of sensor $i$. 
		}
	\label{fig:model:radial_support}
	\if\isextended0\vspace{-0.1in}\fi
\end{figure}

Fig.~\ref{fig:model:radial_support} illustrates examples of sector and omni-directional
radial sensing supports. 
We denote by $S_i = \{X_i + Y_i\} \oplus S_i^0$ the sensing 
support of sensor $i.$ For an object, say $j$, which is not a sensor,
we let $Y_j =0$ and $S_j^0 = \emptyset$.
The environment and the sensing field are thus modeled 
by an Independently Marked PPP (IMPPP), $\tilde{\Phi}$, 
which associates independent marks $M_i=(A_i, Y_i, S_i^0)$ to each object $i$, i.e., 
\begin{equation*}
\tilde{\Phi} = \big\{\big(X_i, M_i \big), i\in {\Natural}^+ \big\}.
\end{equation*}
Note that $(A_i, Y_i, S_i^0)$ is independent of $X_i$, but 
$A_i, Y_i, S_i^0$ need not be mutually independent. 
Indeed if $i$ is a sensor, $Y_i \in A_i$ since the sensor 
should be mounted on the object.
Also the distribution of the shape of objects with sensors, e.g., vehicles, 
can be different from that of other objects, e.g., pedestrians.

\if\isextended1
The aim of such general IMPPP model is to model all the objects in the
environment, including vehicles, pedestrians, motorcycles, buildings, etc., thus
we use a generalized HPPP model for the objects. Note that in practice vehicles
follow the lanes on roads or parking lots, yet the analysis for such settings is
similar to the simplified setting we consider. Furthermore comparisons via
simulation of a detailed freeway model validate that the proposed HPPP model is
a good approximation to study the performance of sensing in typical freeway and
other scenarios. Our model may also apply to other (collaborative) sensing
systems relying on wireless communication, but the model and analysis in this paper
focuses on the unique characteristics of vehicular sensing, i.e., vehicles play
the role of sensor, obstruction, and objects of interest at the same time.
\fi

\subsection{Model for Vehicle's Region of Interest}\label{subsection:region:model}

We shall assume each sensing vehicle 
is interested in information within a certain range 
around it -- usually measured in time, e.g.,  $t_{\interest}$ sec. 
The actual spatial range depends on the vehicle's speed $s$ 
and is given by $s \cdot t_{\interest}$. 
We model a sensing vehicle's \emph{region of interest} as follows.

\begin{definition}(Region of interest)
The {\em region of interest} for sensor vehicle $i$, $D_i$,
is modeled for simplicity as a disc, 
$b(X_i, r)$, centered at $X_i$ with
radius $r = s\cdot t_{\interest}$.
\end{definition}
Note that for a vehicle located at the center of a multi-lane road, 
its region of interest can also be approximated by a rectangular set $[-s \cdot t_{\interest},~~ s \cdot t_{\interest}]
\times [-\frac{w_{\road}}{2}, \frac{w_{\road}}{2}]$, where $w_{\road}$ denotes the 
width of the road. 


\subsection{Collaborative Sensing in an Obstructed Environment}

\begin{figure}[hbt]
	\centering
	\includegraphics[height=0.33\columnwidth]{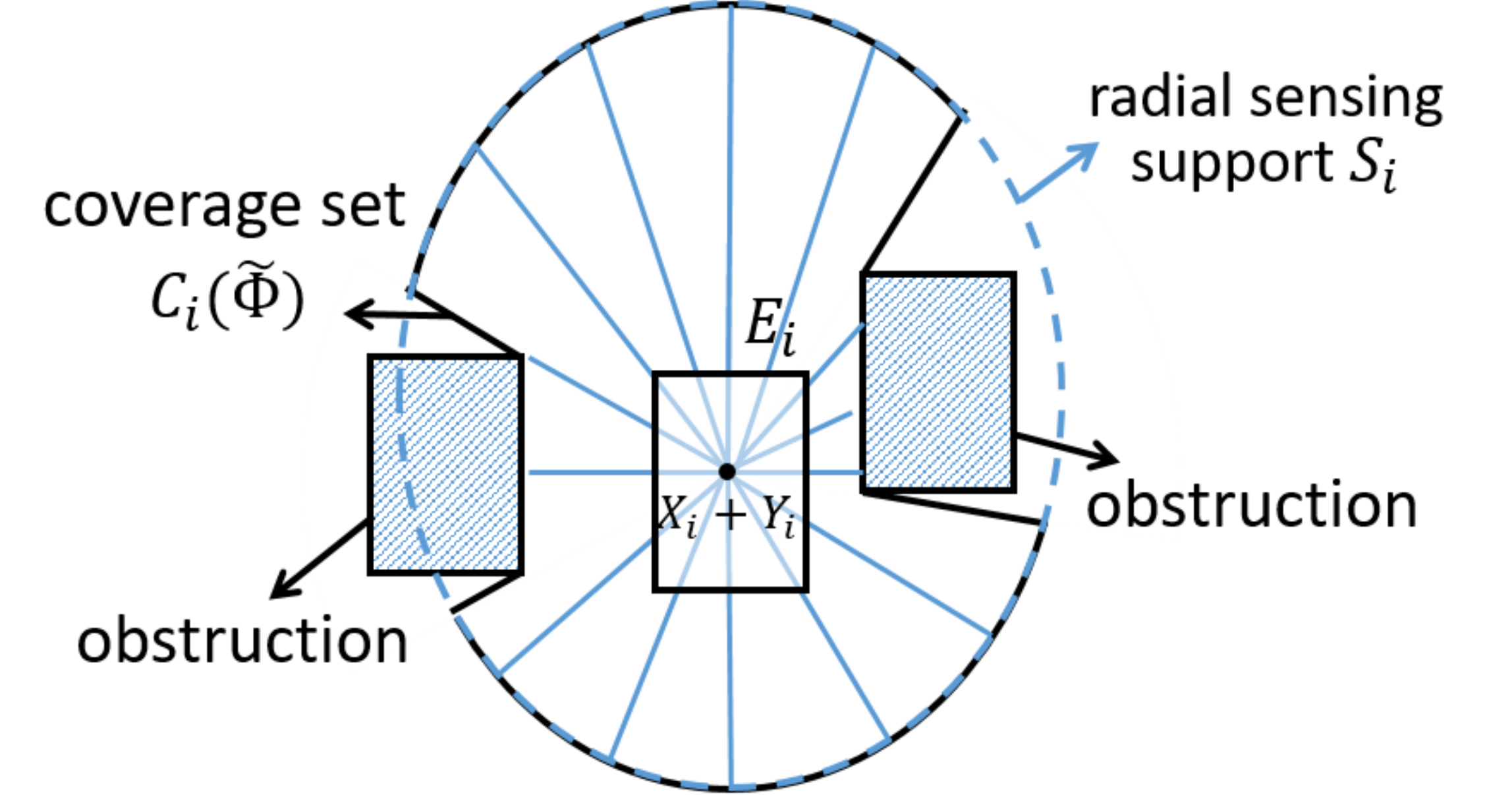}
	\caption[Illustration of sensor $i$ coverage set under blockage]
		{Coverage set of sensor $i$ in $\tilde{\Phi}$.
		}
	\label{fig:model:coverage_set}
	\if\isextended0\vspace{-0.1in}\fi
\end{figure}

Next we define a sensor's coverage set given the environment and sensor model
$\tilde{\Phi}$
as follows -- see Fig.~\ref{fig:model:coverage_set}.

\begin{definition}(Sensor coverage set)
For sensor $i$ with radial sensing support $S_i$ in the environment and sensor
model $\tilde{\Phi}$, we let $E^{-i} = \underset{j: j \neq i }{\cup}E_j$ denote
the environment excluding $E_i$. The {\em coverage set} of sensor $i$,
$C_i(\tilde{\Phi})$, is then given by 
	\begin{equation}\label{eq:model:coverage_set}
		C_i(\tilde{\Phi})  = 
			\big\{x\in S_i\big|x\in E_i \mbox{~or~}l_{X_i+Y_i, x} \cap E^{-i} \subseteq \{x\}\big\},
	\end{equation}
	where $l_{y,z}$ denotes the closed line segment between $y,z \in \Realsquare$. 
The {\em coverage area} of sensor $i$ is the area of 
	its coverage set which we denote $|C_i(\tilde{\Phi})|$. 
\end{definition}

In the above definition, we assume that a sensor is aware of $E_i$, the space it
occupies, i.e. no ``self blocking''.
Also $l_{X_i+Y_i, x}\cap E^{-i}\subseteq \{x\}$ verifies that 
the LOS channel between the sensor at $X_i+Y_i$ and location $x$ is not blocked by other objects.
A location $x\in C_i(\tilde{\Phi})$ may be in the void space or on the surface of an object. 
The coverage set of sensor $i$ represents the surrounding 
environment that it is able to view (on its own) under environmental obstructions.

The expected coverage area of a {\em typical} sensor is given in the 
following theorem, where $C^0$ denotes the coverage set of 
a typical sensor shifted to the origin\footnote{Its distribution is 
formally referred to as the Palm distribution \cite{chiu2013stochastic}.} and 
$A^0$, $Y^0$ and $S^0$ are the associated shape, 
location of sensor, and radial support set, referred to the origin.
The set $\{Y^0\}\oplus S^0 \cap A^0$ denotes the region, if any, 
in the sensing support overlapping with the object, 
while $(\{Y^0\}\oplus S^0)\backslash A^0 = \{x|x\in \{Y^0\}\oplus S^0, x\notin A^0\}$ 
is the region in the sensing support excluding the sensing object. 
Finally $A$ denotes a random set with the same distribution 
as the shape of objects and is independent of $A^0$. Their distributions
may be different, since the latter is conditioned on 
an environmental object being a sensor, i.e., being a sensing vehicle. 

\begin{theorem}\label{theorem:model:coverage_area}
Under our environment and sensor model $\tilde{\Phi}$ 
the expected coverage area of a typical sensor is given by 
\begin{multline}\label{eq:model:coverage_area:expected}
	{\E}\big[|C^0|\big] = \E[|(\{Y^0\}\oplus S^0)\cap A^0|] \\
						 + {\E}\bigg[\int_{(\{Y^0\}\oplus S^0)\backslash A^0} e^{-\lambda\cdot{\E}[|l_{Y^0, x} \oplus \check{A}|]}dx\bigg],
\end{multline}
	where $\check{A} = \{x|-x\in A\}$.
\end{theorem}

For example if objects are modeled as discs of radius $r$, i.e.,
$A = b(0, r)$, with probability $1$, 
and the sensor is mounted at the center, i.e., $Y^0 = 0$, 
we have that $|l_{0, x}\oplus \check{A}| = \pi r^2 + 2r\cdot|x|$ (see \cite{chiu2013stochastic}), 
so $\E\big[|C^0|\big]$ is straightforward to compute.
The theorem shows how the coverage area of a {\em single} 
sensor decreases in the object density $\lambda$ since the probability of sensing
a given location (the term inside integral) decreases exponentially in $\lambda$. 
The proof of Theorem~\ref{theorem:model:coverage_area} leverages straightforward
stochastic geometric results and is relegated to the appendix.

\subsection{Sensor Coverage Area: Numerical and Simulation Results}
\label{subsection:model:coverage_area_simulation}

Below we verify the robustness of our idealized analytical model
by comparing the analytical results to a simulation of vehicles on a freeway. 
For the analytical model, the shape of all objects (vehicles) is a disc of radius $1.67\meter$, 
roughly corresponding to the area of a vehicle, and each has 
an omni-directional sensing support with radius $100\meter$.
For the typical vehicle we limit its sensing support and coverage set to 
a rectangular region of interest centered on the vehicle, say $i$, 
such that \if\isextended0$D_i = b(X_i, 100\meter) \cap ([-\infty, \infty]\times [X_i-12\meter, X_i + 12\meter])$.\else
\begin{equation}\label{eq:model:scenario:analysis:rectangle_region_of_interest}
D_i = b(X_i, 100\meter) \cap ([-\infty, \infty]\times [X_i-12\meter, X_i + 12\meter]).
\end{equation}\fi
This is geared at capturing the fact that vehicles are mainly interested in sensing
nearby road and sidewalks and $12\meter$ is roughly the width of three lanes.

Our simulations are based on the freeway scenario specified in \cite{3gpp5gv2x}
with $3$ lanes in each direction and lane widths of $4$ m.
Vehicles are placed on each lane according to a linear Mat{\'e}rn process \cite{matern2013spatial}, 
i.e., randomly located but ensuring a minimum gap of $10$m among the centers of vehicles on the same lane.
Vehicles are modeled as $4.8$m $\times 1.8$m rectangles, and the distance from 
the center locations to the lane center are uniformly distributed $\unif[-1, 1]$m. 
The coverage area does not include the region off the road.

\if \isextended 1
\begin{figure}
	\centering
	\subfloat[Analytical model]{
		\includegraphics[width=\figurewidth\columnwidth]{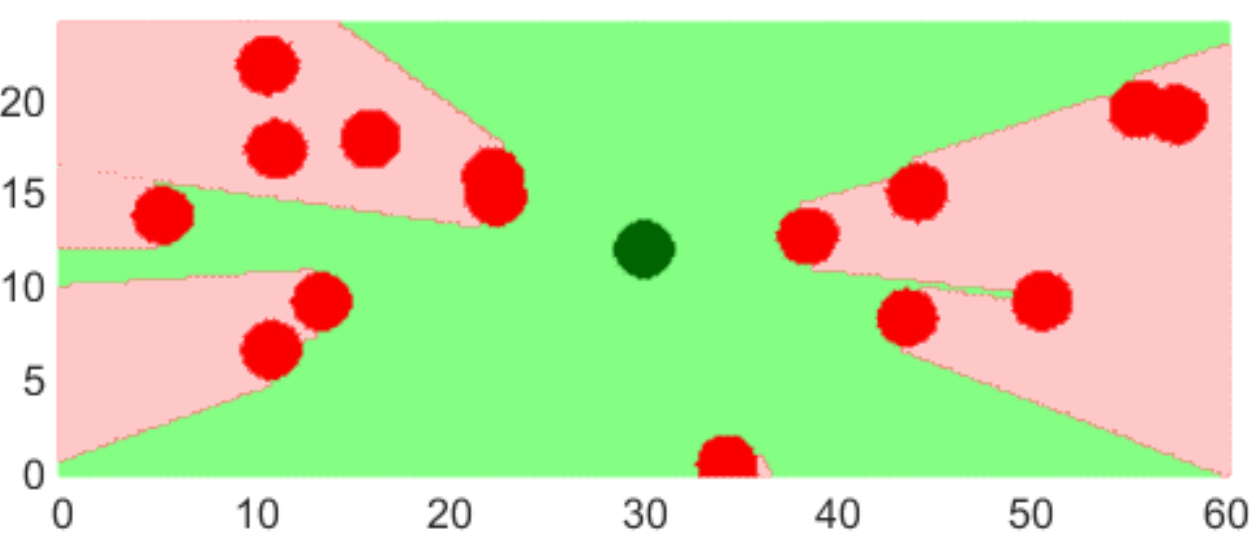}
		\label{subfig:model:sensing:analysis}
	}
	\hfill
	\subfloat[Freeway simulation]{
		\includegraphics[width=\figurewidth\columnwidth]{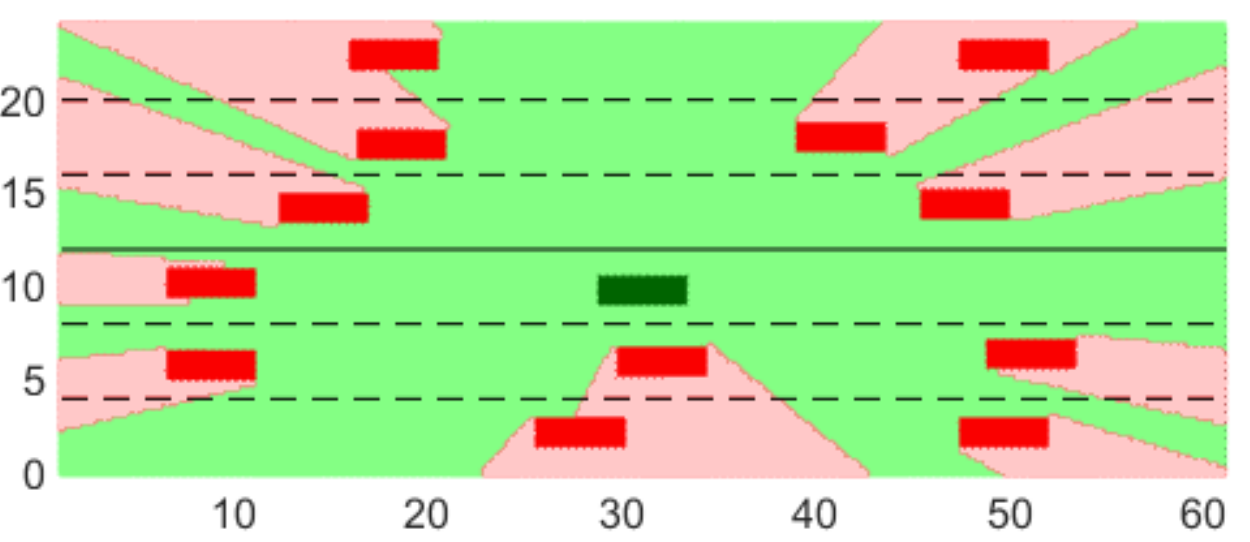}
		\label{subfig:model:sensing:highway}
	}	
	\caption[Sensing in different models]{
		Sensing of a typical vehicle in \subref{subfig:model:sensing:analysis}
		analytical model and \subref{subfig:model:sensing:highway} freeway simulation model.
		The green shapes are reference objects, the red shapes are obstructions,
		light green represents sensed region, light red indicates obstructed region.
	}
	\label{fig:model:scenario}
\end{figure}

Fig.~\ref{fig:model:scenario} illustrates an example of sensing in 
our simplified analytical model and freeway simulation. 
The sensed and obstructed regions in the two models share similar characteristics.
\fi

\begin{figure}
\centering
\includegraphics[width=\figurewidth\columnwidth]{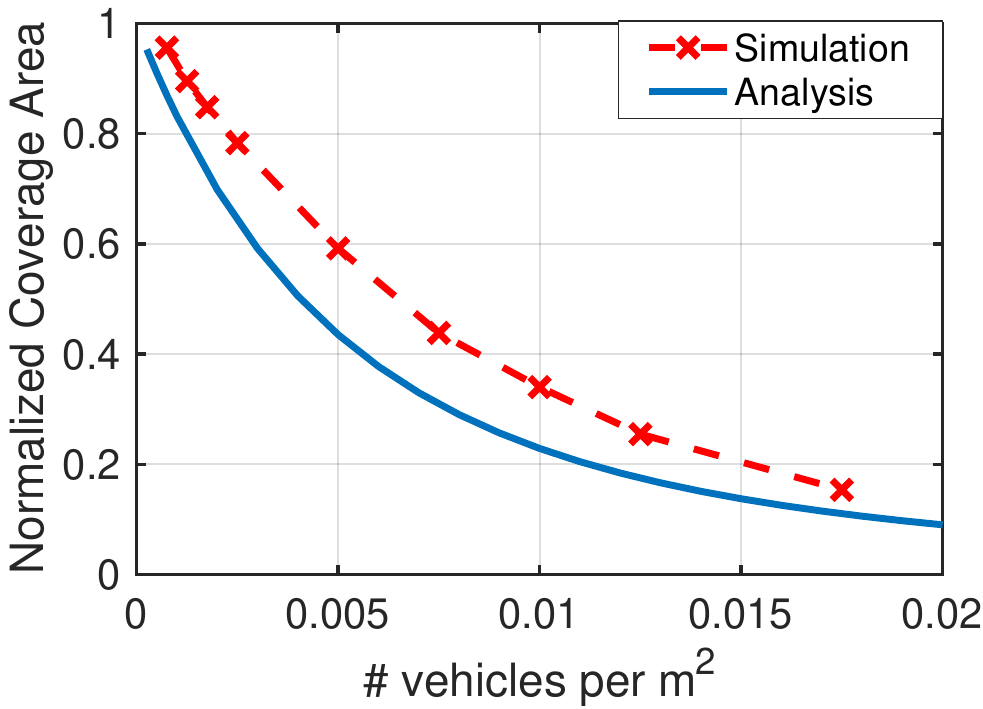}
\caption{Coverage area of a typical vehicle normalized by the area of its sensing support.}
	\label{fig:model:coverage_area}
	\if\isextended0 \vspace{-0.2in}\fi
\end{figure}

Fig.~\ref{fig:model:coverage_area} exhibits analytical and simulation results for 
the vehicle's coverage area normalized by the area
of sensing support scales versus vehicle density $\lambda$. 
Confidence intervals are not shown as they are negligible.
As expected, with increased vehicle density, the sensor coverage area 
decreases due to increased obstructions. 
To reduce boundary effects, the simulation results correspond to 
the average sensor coverage area for vehicles in the two 
most central lanes. 
As can be seen the analytical and simulation results exhibit similar trends. 
At high vehicle densities, the coverage area of a single vehicle 
is heavily limited, i.e., covering less than $20\%$ of the sensing support. 
In an obstructed environment collaborative sensing will be critical to 
achieve better coverage over each vehicle's region of interest. 
We consider this next.

\section{Benefits of Collaborative Sensing}\label{section:collaborative}

The benefits of collaborative sensing are twofold: 
(1) it increases sensing redundancy/diversity leading to improved coverage,
and (2) it improves coverage by effectively extending the sensing range.
We consider two metrics for the performance of collaborative sensing:
\emph{redundancy} and \emph{coverage}. 

\emph{Sensing redundancy.}
We define sensing redundancy as the number of collaborative sensing vehicles 
that can view a location/object. The task of detecting/recognizing and
tracking objects is facilitated if multiple sensors' point of view are available,
providing greater coverage and robustness to sensor/communication link
failures.

\begin{definition}(Sensing redundancy for a location)
Given an environment and sensing field, $\tilde{\Phi}$, 
and a subset of collaborating sensors $K \subseteq \Phi^s$, 
the {\em sensing redundancy for a location $x$} 
is the number of sensors in $K$ that view $x$, denoted by 
	\begin{equation}
	R(\tilde{\Phi}, K, x) = \sum_{i:X_i \in K } \indicator\big(x\in C_i(\tilde{\Phi})\big).
	\end{equation}
\end{definition}
In the most optimistic case $K = \Phi^s$, i.e., all sensors collaborate.
The expected redundancy of a location in the void space is given by the results
in the following theorem. 
\begin{theorem}\label{theorem:collaborative:expected_rendundancy_location}
Given an environment and sensing field $\tilde{\Phi}$ and assuming
all sensors collaborate, $K = \Phi^s$,
the expected redundancy of a typical location $x$ in the void space is 
\begin{equation}\label{eq:model:redundancy:value}
{\E}[R(\tilde{\Phi},\Phi^s, x)|x\notin E] = \frac{p_s\cdot \lambda \cdot{\E}[|C^0\backslash A^0|]}{e^{-\lambda\cdot \E[|A|]}},
\end{equation}
where ${\E}[|C^0\backslash A^0|]$ is given in the second term in
Eq.~\ref{eq:model:coverage_area:expected}.
\end{theorem}
The proof of this theorem included in the appendix follows from the
definition of redundancy and a sensor's coverage set. 

\begin{figure}
	\centering
	\includegraphics[width=\figurewidth\columnwidth]{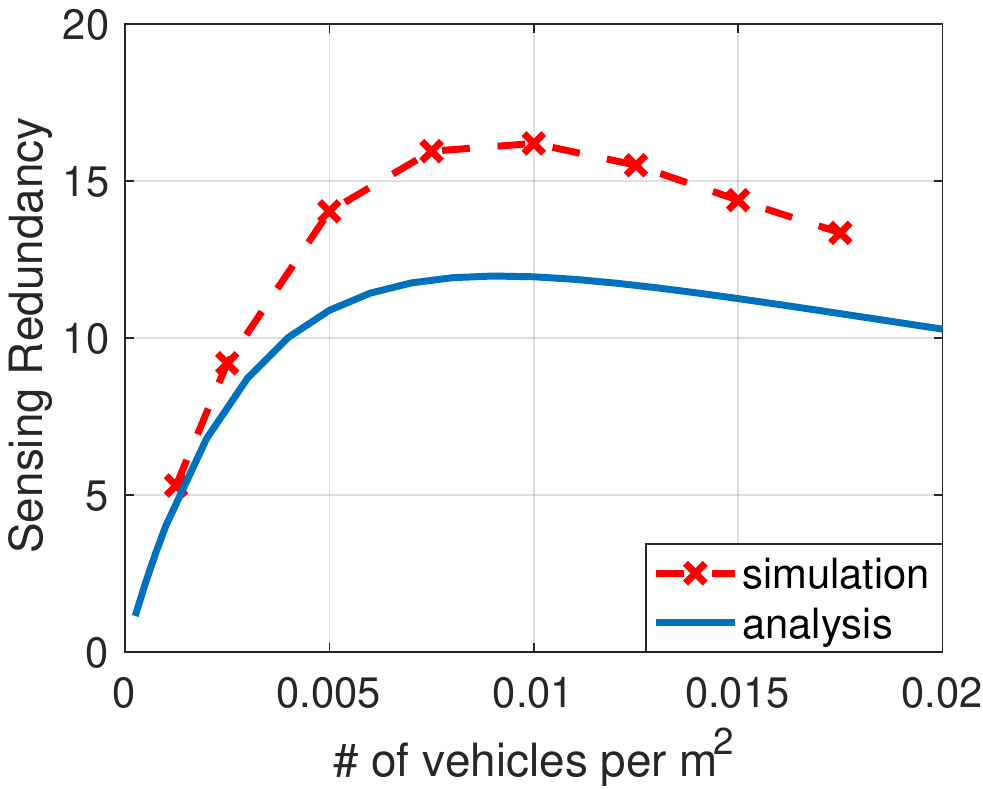}	
	\caption[Redundancy of location at different vehicle densities]
	{The expected sensing redundancy of a random void location versus object density. 
	 All vehicles participate in collaborating, i.e., $K=\Phi^s = \Phi$. }
	\label{fig:collaborative:redundancy_location}
	\if\isextended0 \vspace{-0.2in}\fi
\end{figure}

Fig.~\ref{fig:collaborative:redundancy_location} 
exhibits the expected sensing redundancy of a typical location in the void space.
As can be gleaned from our analytical results, sensing redundancy for
a location is proportional to $p_s$ so we 
only exhibit results for $p_s = 1$. 
At small densities sensors are not likely to be blocked thus
redundancy first increases in the density of objects $\lambda$. 
However, at higher densities, the objects obstruct each other
reducing the coverage area of each sensor and the  
resulting sensing redundancy. 
The simulation results show the expected redundancy of a random 
location 
in the central two lanes
(see Section~\ref{subsection:model:coverage_area_simulation}), 
and exhibit similar trends as the analysis. 
Overall one can conclude that collaborative sensing will
provide highest redundancy at moderate densities, i.e., 
this is where in principle collaborative sensing is most reliable 
and robust to sensor/communication failures. 

\emph{Collaborative sensing coverage.} 
A location in a vehicle's region of interest is covered by 
collaborative sensing if the location can be reliably sensed, i.e., 
sensed by a sufficient number of collaborating sensors. 
We define the collaborative sensing coverage for a vehicle as follows. 


\begin{definition} (Collaborative sensing coverage)
Given an environment and sensing field $\tilde{\Phi}$, 
a minimum redundancy requirement $\gamma \in \Natural^+$ for reliable sensing of a location, 
a subset of collaborating sensors, $K \subseteq \Phi^s$, and sensor $i$'s region of interest $D_i$, 
the {\em $\gamma$-coverage set} of sensor $i$ is the region within $D_i$,
which is covered by at least $\gamma$ sensors in $K$, denoted by 
\begin{equation}\label{eq:model:reliability}
	C_{c}(\tilde{\Phi}, K, D_i, \gamma) \delequal 
			\big\{x \big| x \in D_i, R(\tilde{\Phi}, K, x) \geq \gamma \big\}.
\end{equation}
The \emph{$\gamma$-coverage} of sensor $i$ is the area of the $\gamma$-coverage set, $|C_{c}(\tilde{\Phi}, K, D_i, \gamma)|$. 
The \emph{normalized $\gamma$-coverage} 
is the $\gamma$-coverage normalized by the area of the region of interest,
$|C_{c}(\tilde{\Phi}, K, D_i, \gamma)|/|D_i|$.
\end{definition}

The normalized $\gamma$-coverage can be interpreted 
as the fraction of $i$'s region of interest that can be reliably sensed.
Denote by $D^0$ the possibly random\footnote{Recall the region may depend on the vehicle's speed.}
region of interest associated with a typical sensing vehicle,
and $A^s\subset \Realsquare$ a random set having the same distribution of 
the shape within a the region occupied by the sensor and is covered in the
sensor's support, i.e., $\{Y^0\}\oplus S^0 \cap A^0$.
\if\isextended0
The expected $1$-coverage can then be approximated by
\begin{align}
	\E\big[&|C_c(\tilde{\Phi}, \Phi^s, D^0, 1)|\big] \approx \nonumber\\
	&\E[|D^0\cap C^0|] + \E[|D^0\backslash (C^0\cap A^0)|]\cdot(1 - e^{-\lambda^s\E[|A^s|]}) \nonumber\\
	& +\big(\E[|D^0\backslash A^0|]\cdot e^{-\lambda\E[|A|]} - \E[|D^0 \cap C^0\backslash A^0|]\big)\nonumber\\
	&\cdot(1 - e^{-\E[R(\tilde{\Phi}, \Phi^s, x)|x\notin E]}).
	\label{eq:collaborative:coverage_approximation}
\end{align}
This approximation is based on decomposing $D^0$ into various sets:
$D^0\cap C^0$ the set covered by the sensor, 
and in the region not covered by the sensor, 
we have
$D^0\cap E\backslash C^0$ the set occupied by objects and
$D^0 \backslash (E\cup C^0)$ the void space.
The collaborative coverage in each set can then be evaluated as follows, 
$\E[|D^0 \cap C^0|]$ is area sensed by the vehicle itself,
$\E[|D^0\backslash (C^0\cap A^0)|]\cdot (1 - e^{-\lambda^s\E[|A^s|]})$
the area of region not covered by the sensor but occupied by other sensing vehicle bodies and sensed through collaboration,
$\big(\E[|D^0\backslash A^0|] \cdot e^{-\lambda\E[|A|]} - \E[|D^0 \cap C^0\backslash A^0|]\big)$
is the area of void space not covered by the sensor and
$1 - e^{-\E[R(\tilde{\Phi}, \Phi^s, x)|x\notin E]}$ is an approximation 
for the probability a void location is sensed via collaboration 
which assumes the redundancy of a void location has a Poisson distribution with a mean given in Thm.~\ref{theorem:collaborative:expected_rendundancy_location}.

\fi

\if \isextended 1
\emph{Approximation of the normalized $\gamma$-coverage.}
Denote by $\overbar{Q}(k, m) = \Prob(N(m)\geq k )$, where $N(m)$ is a Poisson random variable with mean $m$.
Denote by $\overbar{R}_{\void} = \E[R(\tilde{\Phi}, \Phi^s, x)|x\notin E]$ 
the expected redundancy of a location in the void space as given in Eq.~\ref{eq:model:redundancy:value}.  
The average $\gamma$-coverage can be approximated by 
\begin{align}
&\E\big[|C_c(\tilde{\Phi}, \Phi^s, D^0, \gamma)|\big] \approx \nonumber\\
&\E[|D^0\cap C^0\cap A^0|]\cdot \overbar{Q}(\gamma - 1, \lambda^s\cdot \E[|A^s|]) \nonumber\\
&+ \E[|D^0\cap C^0\backslash A^0|]\cdot \overbar{Q}(\gamma - 1, \overbar{R}_{\void})\nonumber\\
&+ \E[|D^0\backslash A^0|]\cdot \overbar{Q}(\gamma, \lambda^s\cdot\E[|A^s|]) \nonumber\\
&+ (\E[|D^0 \backslash A^0|]\cdot e^{-\lambda\cdot\E[|A|]} - \E[|D^0\cap C^0\backslash A^0|])\cdot \overbar{Q}(\gamma, \overbar{R}_{\void}).
\label{eq:collaborative:coverage_approximation_gamma}
\end{align}

This approximation is based on decomposing $D^0$ into various sets, 
see Fig.~\ref{fig:collaborative:model_approximation}.
\begin{figure}
	\centering
	\includegraphics[width=\figurewidth\columnwidth]{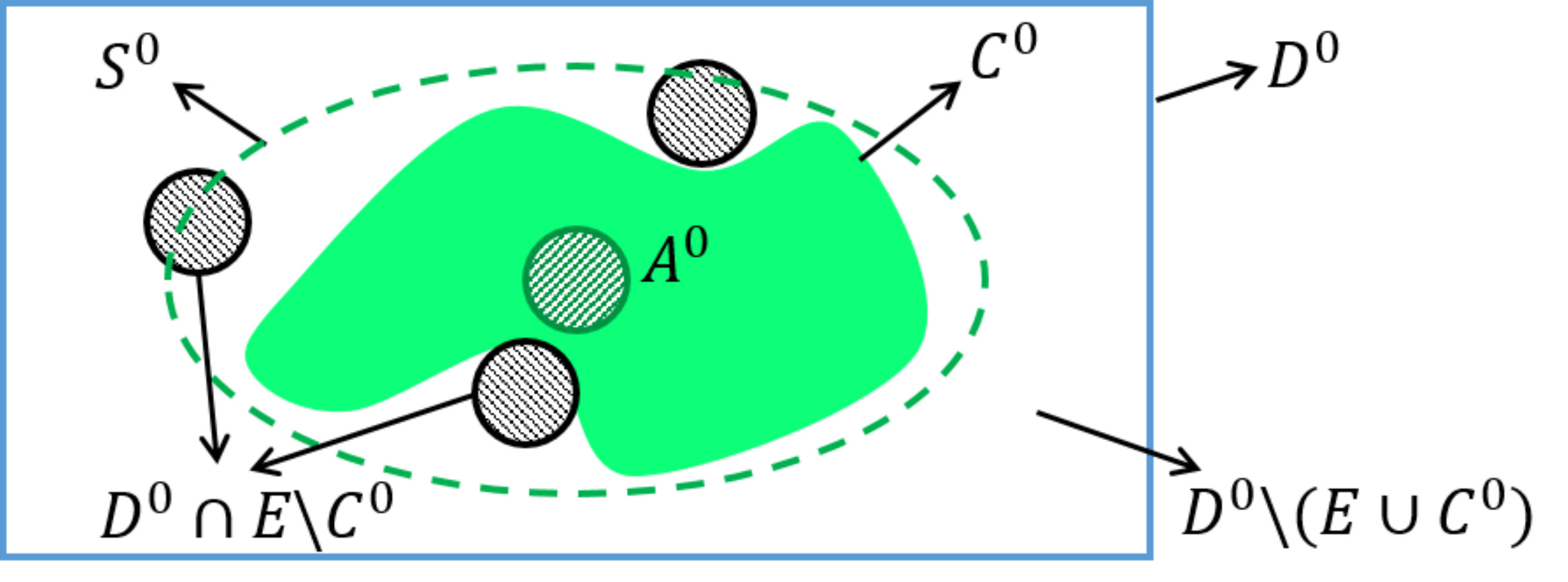}
	\caption{Decomposition of $D^0$ for collaborative sensing coverage approximation.}
	\label{fig:collaborative:model_approximation}
	\vspace{-0.2in}
\end{figure}
In particular, 
$D^0\cap C^0 \cap A^0$ is the set occupied and sensed by the object, 
$D^0\cap C^0 \backslash A^0$ is the set outside the object but sensed by the object,
$D^0\cap E\backslash C^0$ is the set occupied by objects but not in $C^0$, 
and $D^0 \backslash (E\cup C^0)$ is the void space excluding $C^0$.

By Slivynak-Mecke theorem \cite{chiu2013stochastic}, the other objects as seen
by the reference sensor follow an IMPPP with the same distribution as
$\tilde{\Phi}$, so the locations of the other sensors follow HPPP$(\lambda^s)$.
The region covered by objects and sensors will each form a Boolean process \cite{chiu2013stochastic}.
For a random location $x$, 
the number of sensors occupying and sensing $x$ has a Poisson distribution with mean $\lambda^s\cdot\E[|A^s|]$,
the number of objects occupying $x$ has a Poisson distribution with mean $\lambda\cdot\E[|A|]$.
For a location in the void spacer, we will approximate the distribution of the redundancy
by a Poisson distribution with mean $\overbar{R}_{\void}$. 
$\overbar{R}_{\void}$ is not conditioned on there being at typical sensor, 
thus $\overbar{R}_{\void}$ can be different from the expected redundancy at a
location $x\in D^0\backslash E$.
In $C^0$ the reference object provides $1$ redundancy and other sensors 
should provide $(\gamma-1)$ redundancy, while in $D^0\backslash C^0$
the other sensors must provide $\gamma$ redundancy. 

Based on the above approximation, 
the components of Eq.~\ref{eq:collaborative:coverage_approximation_gamma} are interpreted as follows:
\begin{itemize}
	\item $\E[|D^0\cap C^0\cap A^0|]\cdot \overbar{Q}(\gamma - 1, \lambda^s\cdot \E[|A^s|])$ 
	is the area in $A^0$ that is occupied (and sensed) by $\gamma-1$ other sensors.
	\item $\E[|D^0\cap C^0\backslash A^0|]\cdot \overbar{Q}(\gamma - 1, \overbar{R}_{\void})$
	is the area void space space in $C^0$ that is covered by $\gamma - 1$ other sensors. 
	\item $\E[|D^0\backslash A^0|]\cdot \overbar{Q}(\gamma, \lambda^s\cdot\E[|A^s|])$
	is the area in $D^0\backslash A^0$ that is occupied (and sensed) by $\gamma$ sensors.
	\item $\E[|D^0 \backslash A^0|]\cdot e^{-\lambda\cdot\E[|A|]} - \E[|D^0\cap C^0\backslash A^0|]$
	is the area of void space in $D^0\backslash A^0$ excluding $C^0$, 
	and $\overbar{Q}(\gamma, \overbar{R}_{\void})$ is the probability that a location is covered by $\gamma$ other sensors.
\end{itemize}
\fi

\if \isextended 1
\begin{figure}
	\centering
	\subfloat[Sensing in analytical model]{
		\includegraphics[width=\figurewidthnarrow\columnwidth]{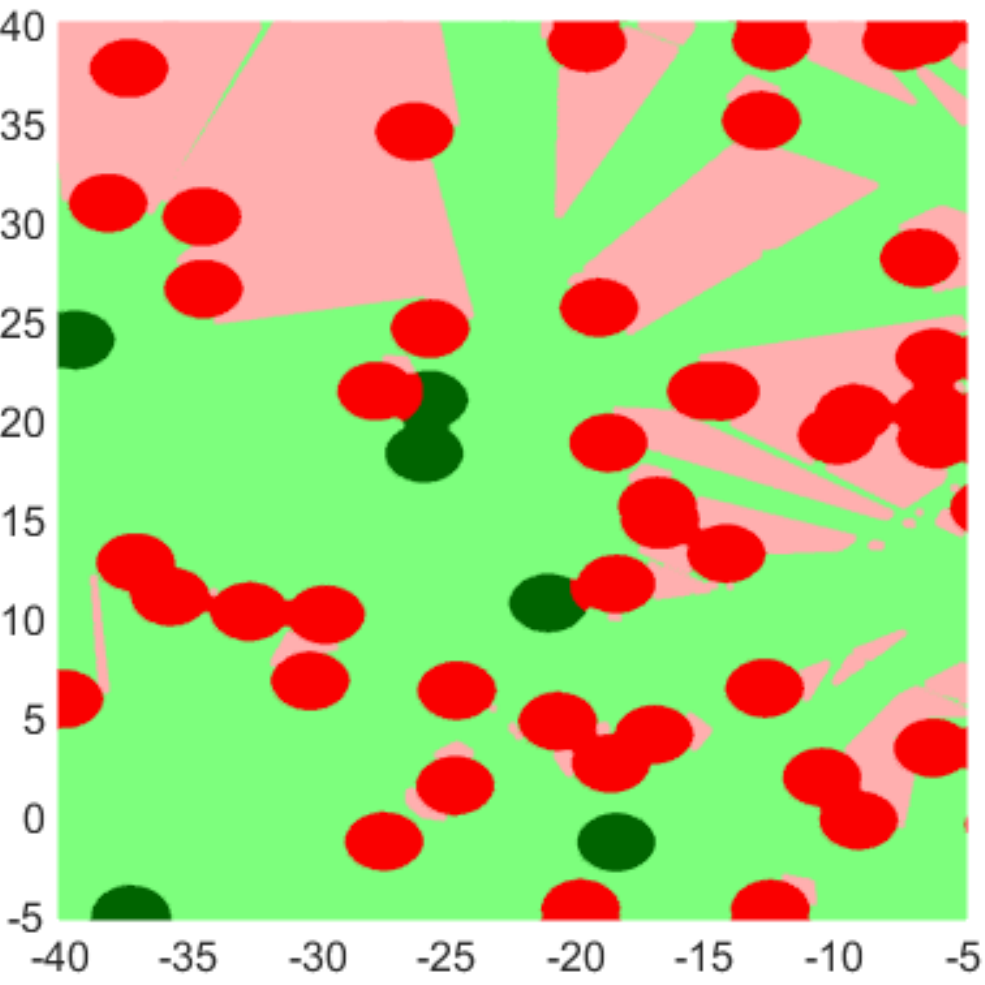}
		\label{subfig:collaborative:scenario:analysis}
	}
	\hfill
	\subfloat[Sensing in freeway scenario]{
		\includegraphics[width=\figurewidth\columnwidth]{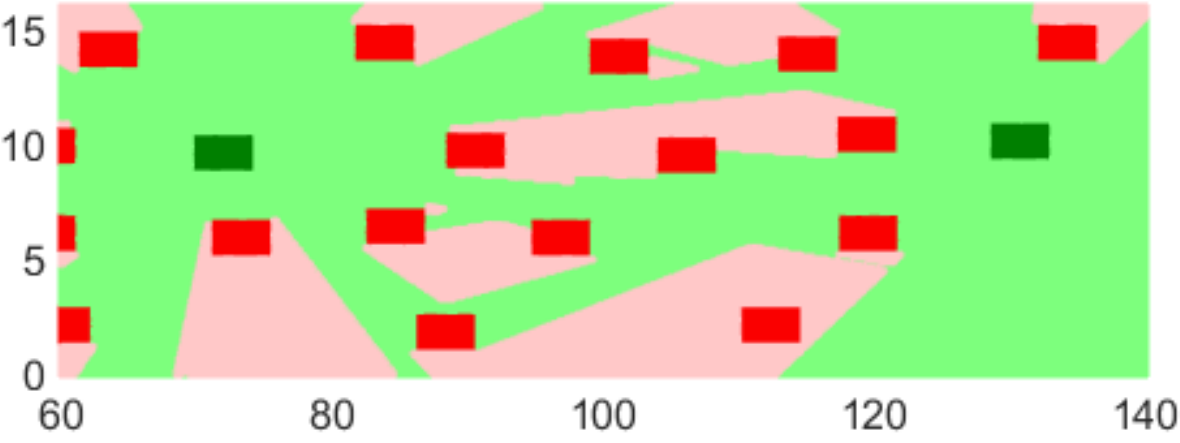}
		\label{subfig:collaborative:scenario:highway}
	}	
	\caption[Collaborative sensing in different models]{
		Collaborative sensing in \subref{subfig:collaborative:scenario:analysis}
		analytical model,
		and \subref{subfig:collaborative:scenario:highway} freeway simulation.
		Dark green shapes represent sensors, red shapes are non-sensing objects. 
		Light green region can be sensed via collaborative sensing, 
		while light red region are obstructed and not sensed. 
	}
	\label{fig:collaborative:scenario}
\end{figure}
Fig.~\ref{fig:collaborative:scenario} illustrates collaborative sensing for our
analytical model and the freeway simulation. 
In the analytical model, objects are modeled as randomly distributed discs 
and may overlap. The objects are randomly placed thus the 
region covered by collaborative sensing will also be the realization of a random
shape. In the freeway simulation, vehicles are randomly distributed along lanes, 
such that there is no overlap. The environment is more `structured', and thus
so is the collaborative sensing coverage set, e.g., the space between lanes
is less likely to be obstructed.
\fi

\begin{figure}
	\centering
	\subfloat[Approximation validation]{
		\includegraphics[width=\figurewidth\columnwidth]{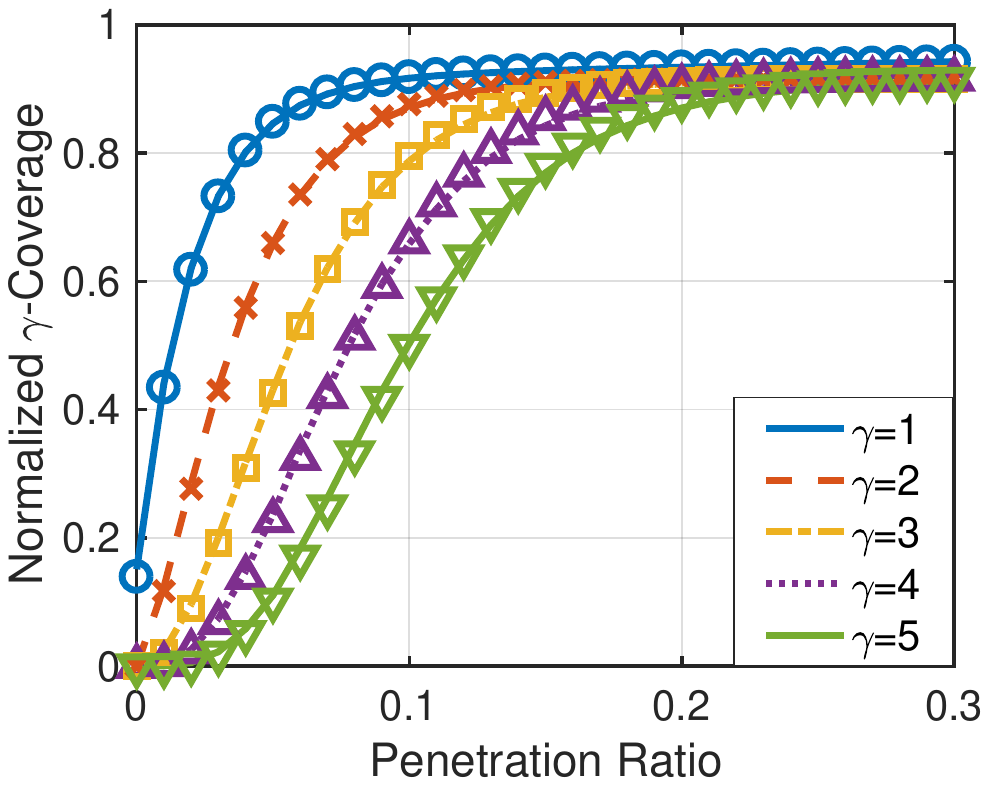}
		\label{subfig:collaborative:reliability:validation}
	}
	\hfill
	\subfloat[Freeway simulation model]{
		\includegraphics[width=\figurewidth\columnwidth]{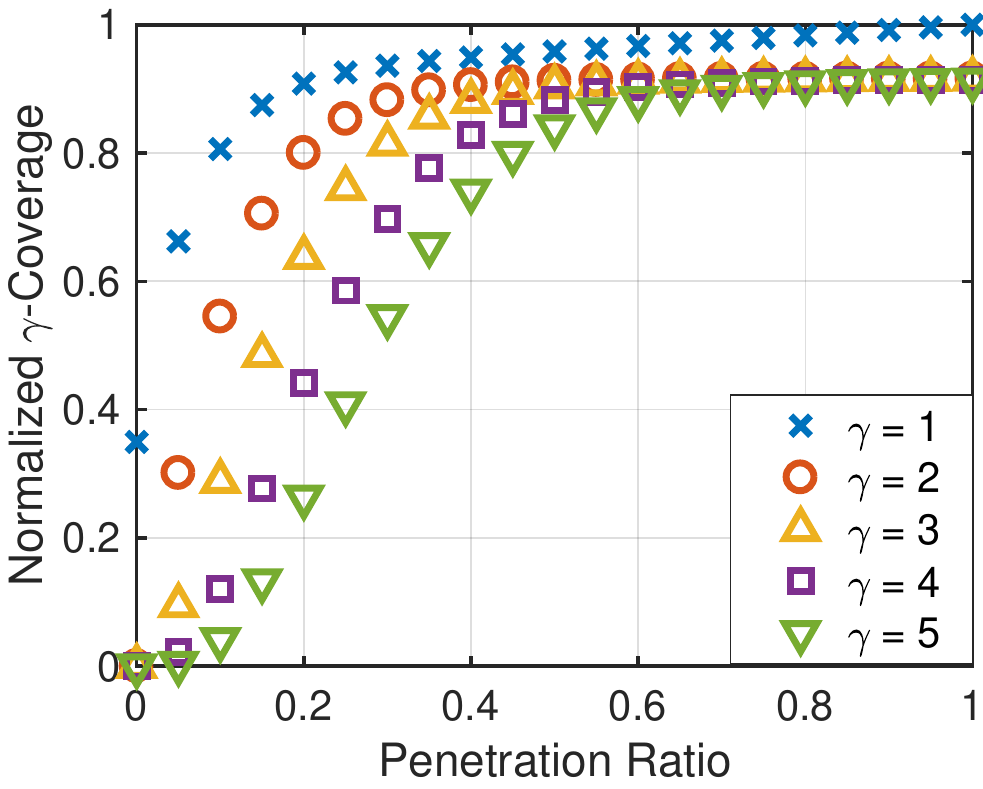}
		\label{subfig:collaborative:reliability:simulation_gamma}
	}	
	\caption[Coverage for different redundancy requirement]{
		Normalized $\gamma$-coverage for different redundancy requirements $\gamma$.
		In \subref{subfig:collaborative:reliability:validation} curves represent results from our approximation\if\isextended 1
		in Eq.~\ref{eq:collaborative:coverage_approximation_gamma}	
		\fi, markers are simulation of the analytical model. \subref{subfig:collaborative:reliability:simulation_gamma} are freeway simulation results.
	}
	\label{fig:collaborative:reliability:different_redundancy_threshold}
\end{figure} 

We validate the accuracy of our approximation in
Fig.~\ref{subfig:collaborative:reliability:validation},
in which we consider a 2D infinite plane with $\lambda = 0.01/{\meter}^2$.
As can be seen the approximation in Eq.~\ref{eq:collaborative:coverage_approximation_gamma}
is a good match for the analytical model.
Fig.~\ref{subfig:collaborative:reliability:simulation_gamma} exhibits 
the freeway simulation results, which show the same qualitative 
trend as the analytical results.
As expected the minimum penetration to achieve a certain level of $\gamma$-coverage
increases in the required diversity $\gamma$.

\begin{figure}
	\centering
	\subfloat[Analytical Approximation]
	{\label{subfig:collaborative:reliability:analysis}
		\includegraphics[width=\figurewidth\columnwidth]{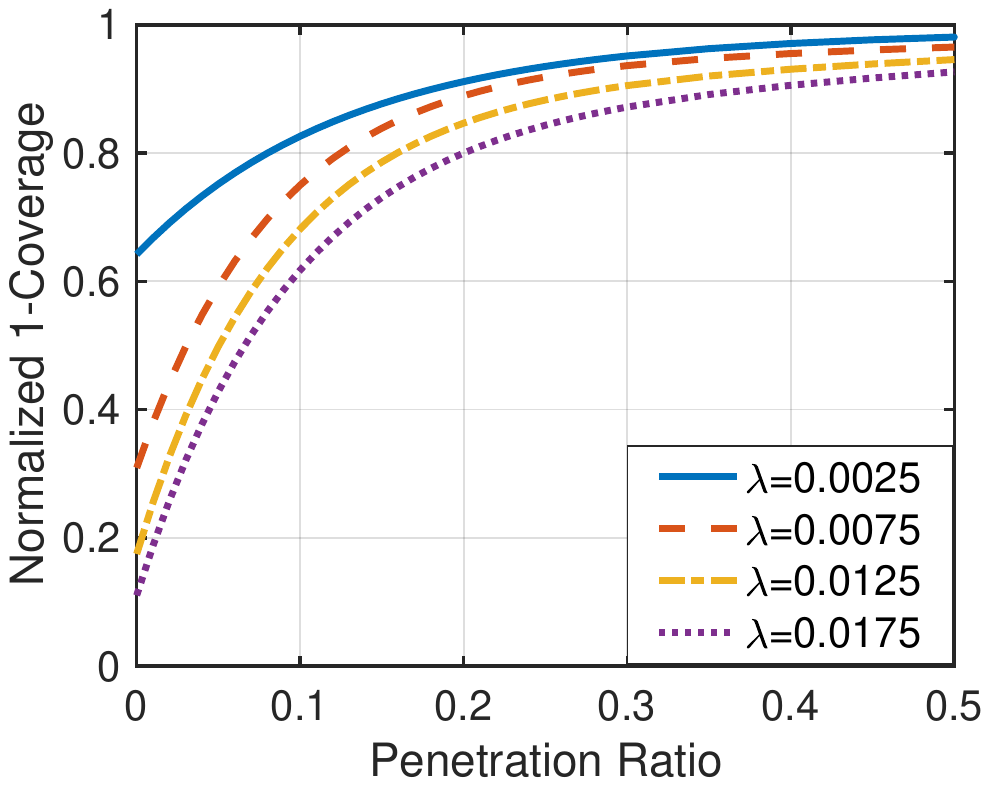}} 
	\hfill
	\subfloat[Freeway simulation model]
	{\label{subfig:collaborative:reliability:simulation}	
		\includegraphics[width=\figurewidth\columnwidth]{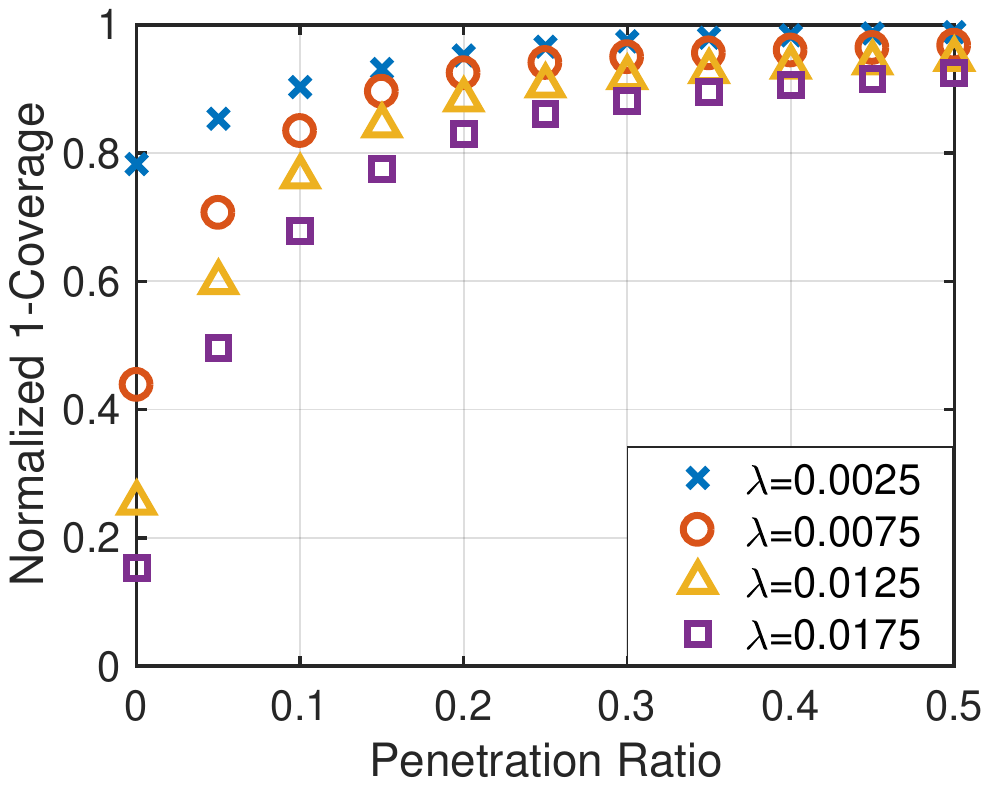}}
	\caption[Collaborative sensing reliability at different penetration of collaborative vehicles]
	{	Normalized $1$-coverage:
		\subref{subfig:collaborative:reliability:analysis} based on analytical approximation in \if\isextended0 Eq.~\ref{eq:collaborative:coverage_approximation},
		\else Eq.~\ref{eq:collaborative:coverage_approximation_gamma}, \fi and 
		\subref{subfig:collaborative:reliability:simulation} obtained by simulation of freeway scenario. 
	}
	\label{fig:collaborative:reliability}
\end{figure} 

Fig.~\ref{fig:collaborative:reliability} exhibits the expected normalized $1$-coverage 
for varying penetrations $p_s$ and vehicle densities $\lambda$. 
The freeway simulation results show the same trend as analytical results. 
Note that \if\isextended0 
Eq.~\ref{eq:collaborative:coverage_approximation}
\else 
Eq.~\ref{eq:collaborative:coverage_approximation_gamma}
\fi 
is an approximation of the analytical model, which is different from 
our simulation of the freeway scenario.
As expected, coverage increases monotonically in $p_s$. 
More importantly, collaborative sensing can greatly improve coverage
even with a small penetration of collaborating vehicles, e.g., 
over $0.8$ coverage when $20\%$ of vehicles collaborate as compared to 
$0.2$ coverage without collaboration at a vehicle density $\lambda=0.0175/{\meter}^2$. 
Such results indicate that it would be beneficial to share sensor data
even with only a subset of neighboring vehicles.

Despite the performance gains associated with vehicular collaborative sensing, 
achieving a high $\gamma$-coverage at low penetrations is 
difficult, especially for $\gamma > 1$.
Joint collaborative sensing with Road Side Units (RSUs) having sensing capabilities
can help improve coverage, e.g., and RSU infrastructure could provide
$100\%$ $1$-coverage if located above a freeway 
(no obstruction) if their sensing support covers the freeway.
If $\gamma_{\rsu}$ denotes the redundancy provided by RSU infrastructure,
the {\em gain} in $\gamma$-coverage associated with joint vehicle/RSU collaboration 
is given by
\begin{equation}
\E\big[|C_c(\tilde{\Phi}, \Phi^s, D^0, \gamma-\gamma_{\rsu})|\big] - \E\big[|C_c(\tilde{\Phi}, \Phi^s, D^0, \gamma)|\big].
\end{equation}
In Fig.~\ref{subfig:collaborative:reliability:simulation_gamma}, 
for $\gamma_{\rsu} = 1$ and $p_s = 0.1$, collaboration with RSUs improves
$2$-coverage by over $0.25$.
In summary the possibility of combining vehicular collaborative sensing with
infrastructure (RSU) based sensing provides a natural avenue to improve coverage,
especially at low penetrations, but possibly also at higher penetrations if 
$\gamma=2$ or higher diversity is desired.


\if \isextended 1 
Another setting our analytical framework can shed light on is how the 
collaborative sensing coverage scales in the obstruction density 
when the sensor density is fixed. 
One example of such a setting might be some freeway on ramps 
where vehicles entering the freeway are primarily non-sensing capable. 
Fig.~\ref{fig:collaborative:coverage_fixed_sensor_density} exhibits how the
$1$-coverage scales in the obstruction density based on the
approximation in Eq.~\ref{eq:collaborative:coverage_approximation_gamma}.
As can be seen the $1$-coverage decreases approximately linearly in the
obstruction density. Collaborative sensing with RSUs may be required to
ensure coverage in such scenarios. 

\begin{figure}
	\centering
	\includegraphics[width=\figurewidth\columnwidth]{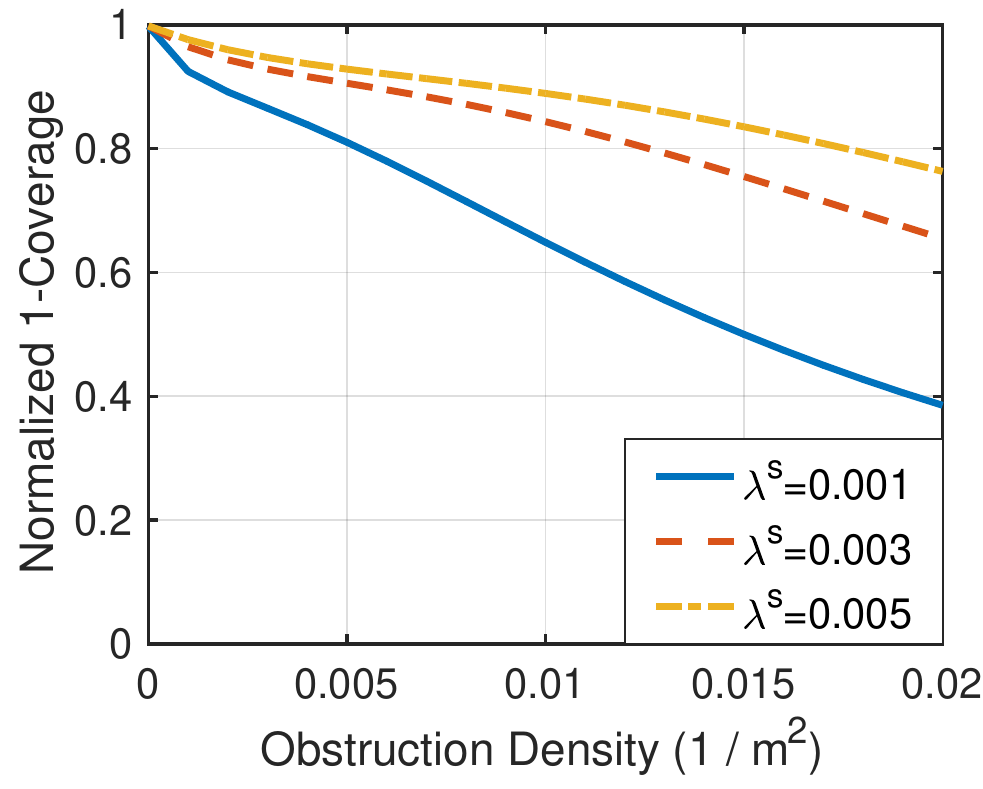}
	\caption{Normalized $1$-coverage for different obstruction densities, $\lambda-\lambda^s$. Sensor density $\lambda^s$ is fixed. }
	\label{fig:collaborative:coverage_fixed_sensor_density}
\end{figure}
\fi

\section{Network Capacity Scaling for Collaborative Sensing Applications}
\label{section:scalability}
In this section we study the network capacity requirements 
for collaborative sensing. 
We envisage both V2V and V2I connectivity might be used to
enable collaborative sensing in automotive settings.
This might be critical to meet reliability and
coverage requirements as we transition from legacy systems.
In particular when the penetration of collaborative sensing vehicles is limited, 
the V2V links/paths required to share collaborative sensing data may be blocked
/ unavailable, particularly when line of sight based links are used such as
mmWave or optical based links.
When this is the case, V2I connectivity, e.g., LTE based links, could serve as
the fallback to share critical sensing/manouvering information.
Below we study the V2I fallback capacity scaling for collaborative sensing settings.

\begin{figure}
	\centering
	\includegraphics[width=0.8\columnwidth]{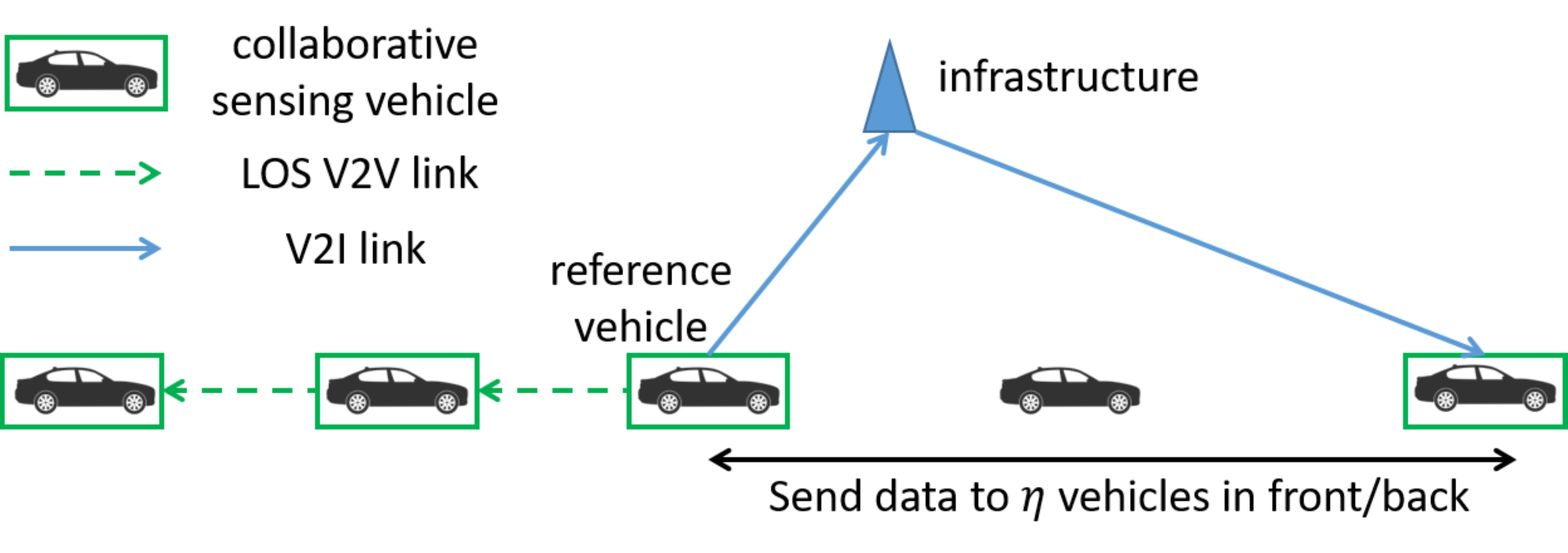}
	\caption[Collaborative sensing of vehicles in a single lane with V2V + V2I network.]
	{Collaborative sensing of vehicles in a single lane with V2V + V2I network. Vehicle uses V2I to
		relay data when LOS V2V links are blocked.}
	\label{fig:scalability:v2i_1d_example}
	\if\isextended0 \vspace{-0.2in}\fi
\end{figure}

We consider vehicles on a single lane assisted by infrastructure deployed along the road.
Vehicles move at a constant velocity $s$.
Each sensing vehicle has a region of interest, $t_{\interest}$ sec, in both 
forward and backward directions.
We shall consider the worst case scenario, i.e., the density of vehicles is high and the gap between (the centers of) vehicles in the
same lane is the minimum gap for safe driving, $t_{\gap}$ sec and the 
inter-vehicle gap is $s\cdot t_{\gap} \meter$. The density of vehicle is thus
given by $\lambda_{\vehicle} = \frac{1}{s\cdot t_{\gap}}$.
We assume vehicles need to receive data from all vehicles in their range
of interest, and by symmetry vehicle also need to send data to all vehicles
in their range of interest.
A sensing vehicle thus needs to send data to $\eta = \lfloor \frac{t_{\interest}}{t_{\gap}}\rfloor$ 
other vehicles in front and behind it, see Fig.~\ref{fig:scalability:v2i_1d_example}.

A vehicle has LOS V2V communication channels to the neighboring vehicles in front and back. 
A non collaborating vehicle thus blocks the V2V relay path along the chain of vehicles.
If a LOS V2V relay path is not available, we assume the reference vehicle relays data
through the infrastructure and the receiving vehicle
can then further relay data to other vehicles via available V2V
links (V2I + V2V relay).
We assume the message a vehicle sends to other vehicles is the same, thus a vehicle
only needs to upload its data to the infrastructure at most once. The
infrastructure can then relay the message to other vehicles requiring the message
via either unicast or broadcast. If using unicast, infrastructure needs to send
the message to every vehicle located in its service region, which requires the
message and cannot get the message via V2V / V2I + V2V relay.

Let $N_{\uplink}$ and $N_{\downlink}^{\unicast}$ be random variables 
denoting the number of uplink and unicast downlink V2I transmissions required to
share data of a typical sensing vehicle.
The expected required V2I uplink capacity $c_{\uplink}$ and 
V2I downlink capacity for broadcast, $c_{\downlink}^{\broadcast}$, and unicast, 
$c_{\downlink}^{\unicast}$, are given in the following theorem. 
\begin{theorem}\label{theorem:scalability:v2i:1d_capacity}
	Consider a single lane model, with a density of vehicles is $\lambda_{\vehicle}$, 
	where each sensing vehicle share data with
	$\eta = \lfloor t_{\interest}/t_{\gap}\rfloor$ vehicles in front and back.
	The V2I capacity requirements on a infrastructure serving the linear road
	segment of length $d\meter$ are given by
	\begin{align}
	c_{\uplink} = c_{\downlink}^{\broadcast} & = p_s\cdot \lambda_{\vehicle}\cdot d\cdot\E[N_{\uplink}]\cdot \nu, \\
	c_{\downlink}^{\unicast} & = p_s\cdot \lambda_{\vehicle}\cdot d\cdot\E[N_{\downlink}^{\unicast}]\cdot\nu, 
	\end{align}
	where
	\begin{equation}
		\E[N_{\uplink}] = 1 - \big(\sum_{k = 0}^{\eta} p_s^{k}\cdot(1-p_s)^{\eta - k}\big)^2, 
	\end{equation}
	\begin{equation}\label{eq:scalability:v2i:1d:n_dl_u}
	\E[N_{\downlink}^{\unicast}]=
	\begin{cases}
	2(\eta-1)p_s(1-p_s),& \text{if } \eta\geq 2,\\
	0,              & \text{otherwise}
	\end{cases}.
	\end{equation}
\end{theorem}
The development of this result can be found in the appendix. The above results
convey the {\em average} capacity requirements on V2I infrastructure.
Unfortunately in a single lane setting a single non-collaborating vehicle can
block the V2V LOS links/paths amongst a large number of vehicles and result in
a {\em burst} of V2I traffic especially at high penetrations, e.g., when
vehicles in front and back of the non-collaborating vehicle are all
collaborating. The required V2I capacity to handle such bursts can thus be much
higher. 

\begin{figure}
	\centering
	\includegraphics[width=0.8\columnwidth]{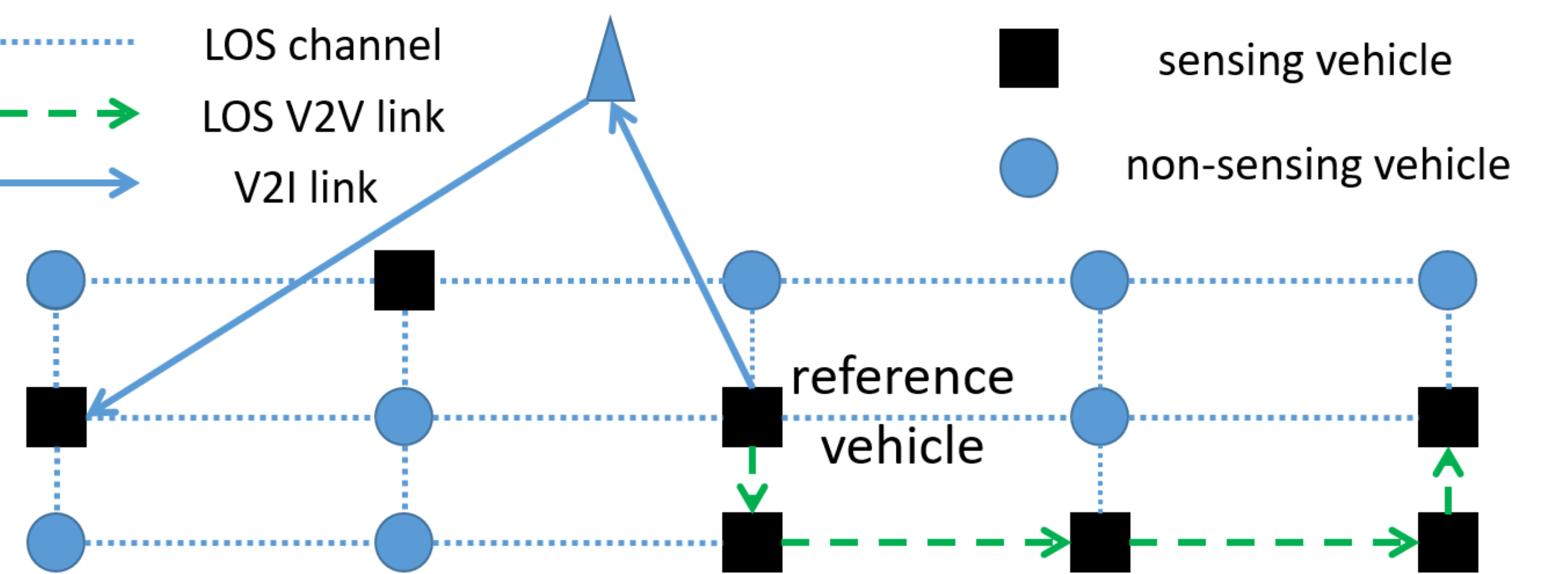}
	\caption[Collaborative sensing of vehicles in a single lane with assistance from vehicles in the two neighboring lanes]
	{Collaborative sensing of vehicles in a single lane using V2V + V2I, with V2V relay assistance from vehicles in the two neighboring lanes.}
	\label{fig:scalability:v2i_2d_model}
	\if\isextended0 \vspace{-0.2in}\fi
\end{figure}

The single lane relaying scenario studied above is a worst case,
i.e., data can only be relayed by vehicles on the same lane.
One can also consider scenarios where
in addition collaborative vehicles on
either of two neighboring lanes participate in V2V relaying.
LOS links among vehicles on neighboring lanes are less likely to be blocked,
but LOS links to distant
vehicles in neighboring lanes will see larger path loss
and may experience more interference, e.g.,
from transmissions of vehicles in the same lane.
Thus for simplicity suppose vehicles only communicate with
the closest vehicle in a neighboring lane and consider the simple grid connectivity
model shown in Fig.~\ref{fig:scalability:v2i_2d_model}.
Each node on the grid corresponds to a vehicle, and each row represents a lane.
Vehicles have LOS channels to neighboring vehicles on the grid.
For comparison purposes we suppose, as before, that a reference vehicle needs
to send data to $\eta$ vehicles in front and back in the {\em same} lane.
Vehicles can receive data via V2V links if there is an LOS V2V
relay path on the grid. To limit the number of hops and associated delays
we assume that a relay path can not include links in {\em both} 
forward and backward directions.

Based on this model, whether vehicles in the $(k+1)^{th}$ column
from the reference vehicle can receive data via V2V links
depends on whether the vehicles in the $(k+1)^{th}$ column are collaborating
and can get data from vehicles in the $k^{th}$ column.
In this setting one can again compute the expected 
V2I capacity requirements to deliver data to vehicles in each column 
and thus the total capacity requirements  
as a function of $\eta$ and $p_s$
-- a detailed analysis is included in \if\isextended0 the extended version of this paper \cite{wang2017scalabilityextended}.\else the appendix.\fi

\subsection{Numerical Results}
\begin{figure}
	\centering
	\subfloat[Single lane]{
		\includegraphics[width=\figurewidth\columnwidth]{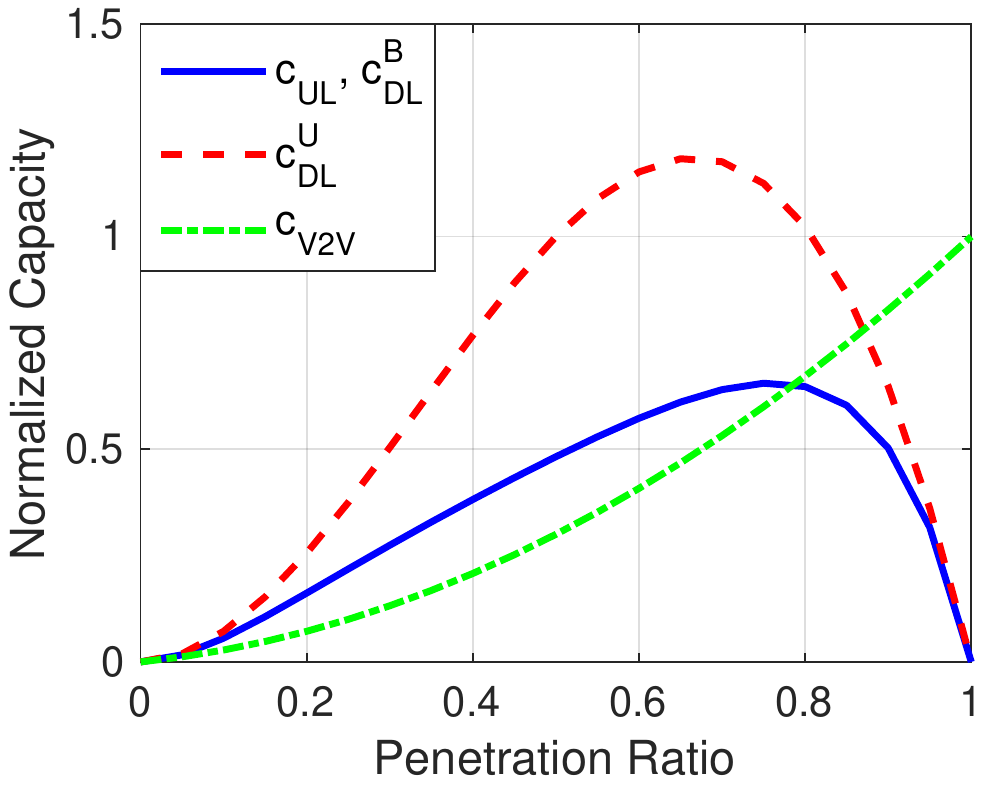}
		\label{subfig:scalability:v2i:1d:capacity}
	}
	\hfill
	\subfloat[Single lane assisted by vehicles in neighboring lanes]{
		\includegraphics[width=\figurewidth\columnwidth]{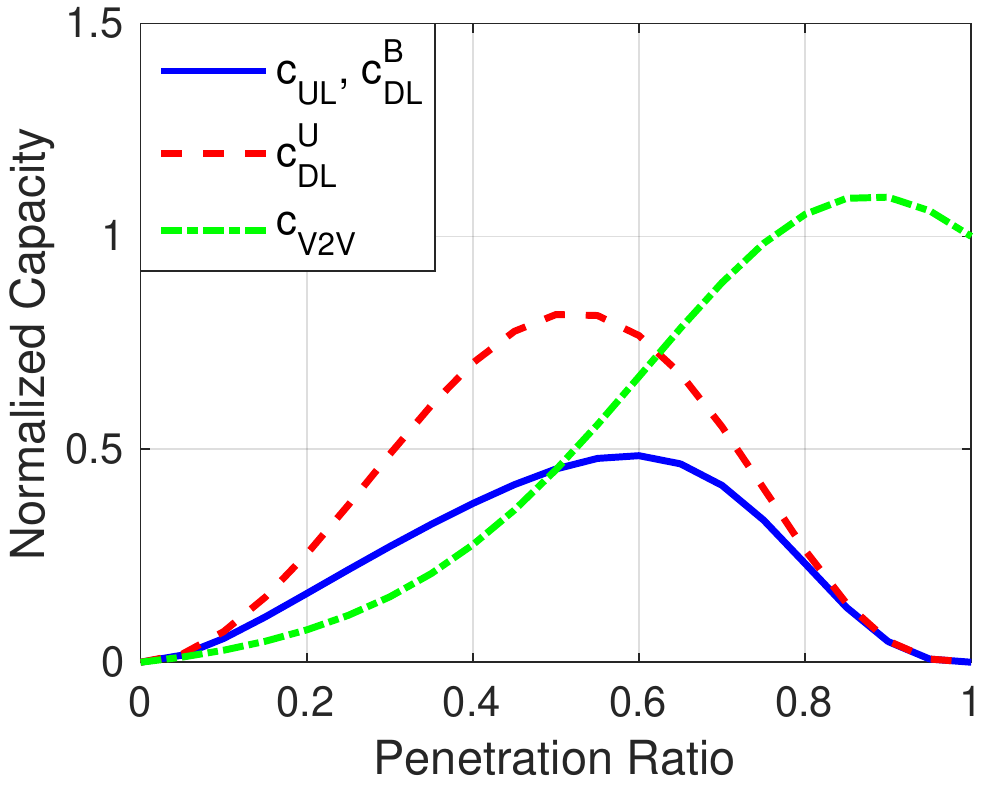}
		\label{subfig:scalability:v2i:2d:capacity}
	}

	\caption[Scalability of V2I capacity]{
		How V2I capacity requirements, normalized by $\lambda_{\vehicle}\cdot d \cdot \nu$, 
		scale with $p_s$ in \subref{subfig:scalability:v2i:1d:capacity} single lane and \subref{subfig:scalability:v2i:2d:capacity} single lane assisted by vehicles in neighbor lanes.
		$c_{\uplink}$ is uplink capacity, 
		$c_{\downlink}^{\broadcast}$ and $c_{\downlink}^{\unicast}$ are downlink capacity using broadcast and unicast.
		$c_{\vtov}$ is V2V throughput per sensing vehicle normalized by the V2V throughput at full penetration, $p_s = 1$.
	} 
	\label{fig:scalability:v2i_capacity}
\end{figure}

Fig.~\ref{fig:scalability:v2i_capacity} exhibits how the
V2I capacity, $c_{\uplink}$, $c_{\downlink}^{\broadcast}$ and $c_{\downlink}^{\unicast}$,
normalized by $\lambda_{\vehicle}\cdot d \cdot \nu$
and the average V2V throughput per sensing vehicle 
normalized by V2V throughput at $p_s=1$, 
vary with $p_s$ in single lane setting and in the single lane assisted by
vehicles in neighboring lanes setting.
The results correspond to the case where $\eta = 5$. 
An increase in $p_s$ causes an increase in the number of vehicles 
participating in collaborative sensing but also results in improved V2V connectivity. 
When $p_s$ is small, both the number of collaborative sensing vehicles 
and the capacity per sensing vehicle increase, thus V2I traffic increases. 
However at higher penetrations, V2V connectivity improves and 
the V2I capacity requirements of a sensing vehicle decreases, 
resulting in lower and eventually negligible V2I traffic.
Comparing the results with and without assistance from vehicles in neighboring lanes, 
we observe, as expected, that V2I traffic is smaller when vehicles in neighboring lanes can help relay data. 
The V2V throughput per sensing vehicle increases with $p_s$. 
However if vehicles in neighboring lanes assist with V2V relaying, the V2V 
throughput is higher than that in the single lane scenario, and
the $c_{\vtov}$ can be higher than the V2V throughput at full penetration.

In summary the V2I traffic resulting from collaborative sensing data would be 
highest at intermediate penetrations, e.g., ranging from $0.5$ to $0.7$,
but eventually would decline once most vehicles participate in both 
collaborative sensing and V2V networking. This suggests an evolution path where V2I resources 
are initially critical to safety-related services like collaborative sensing, 
but eventually at high penetrations of sensing vehicles, 
traffic can be effectively offloaded to the V2V network, e.g., in the single
lane assisted by neighboring lanes, 
$c_{\uplink},c_{\downlink}^{\broadcast}$ per vehicle is less than $0.25\nu$ if $p_s>0.8$,
and the infrastructure may transition to supporting non-safety-related services, 
e.g., mobile high data rate entertainment and dynamic digital map updates.
These results are likely robust to improved models, yet more detailed 
analysis based on more accurate V2V mmWave 
channel and networking models would be required to provide more accurate 
quantitative assessment.

\if\istvt1
\input{sensing_7_dynamics_shorten.tex}
\else
\section{Impact of Dynamics on Collaborative Sensing}
\label{section:dynamics}

In the previous sections we studied how collaborative sensing improves 
coverage for a \emph{snapshot} of the environment by providing spatial 
diversity in sensing, i.e., sensor data for locations and objects 
from different points of view.
In addition, collaborative sensing can improve sensing performance 
by utilizing temporal diversity in sensing. 
Objects in the environment are moving thus the environment is 
dynamic, e.g., vehicles' regions of interest, blockage fields, and 
the sensor coverage sets are varying with time.
Sensor data measured at different time provides possibly different 
information regarding the environment, thus sensors can exploit
temporal diversity for sensing and tracking of objects in the environment. 

\subsection{Temporal Dynamic Environment and Sensing Model}
We shall consider extending the environment and sensing model proposed in
Section~\ref{section:model} to capture temporal dynamics. 
We let $X_i$ be the location of object $i$ at time $0$, and 
denote by 
\begin{equation*}
\Phi^d(t) = \{X_i^d(t), i\in {\Natural}^+\}
\end{equation*}
the locations of objects at time $t$, where $X_i^d(t)$ is the 
location of object $i$ at time $t$. 
Suppose the movements of objects are IID and independent 
of the locations of objects (during the time interval of interest).
Since the objects' locations follow $\Phi\sim \HPPP(\mathrm{\lambda})$ it
follows by the Displacement Theorem \cite{chiu2013stochastic} that the locations
of objects at any time $t$ will remain an HPPP process, i.e., for all $t>0$, 
$$\Phi^d(t)\sim \HPPP(\lambda).$$ 

For simplicity we suppose the shape, location of sensor on the object, 
and the radial sensing support of the sensor, $M_i = (A_i, Y_i, S_i^0)$, 
do not change with time, e.g., objects do not rotate. 
Denote by 
\begin{equation}
E_i^d(t) = X_i^d(t) \oplus A_i
\end{equation} 
the region occupied by object $i$ at time $t$, and 
\begin{equation}
S_i^d(t) = X_i^d(t)\oplus S_i^0
\end{equation}
the sensing support of sensor $i$ at $t$.
The environment and sensing sensing field at $t$ is then given by 
\begin{equation*}
\tilde{\Phi}^d(t) = \big\{\big(X_i^d(t), (A_i,Y_i,S_i^0)\big), i\in {\Natural}^+\big\},
\end{equation*}
and the model for the temporal dynamics of the environment and 
sensing capabilities is denoted by
\begin{equation*}
\boldsymbol{\tilde{\Phi}^d} = \big(\tilde{\Phi}^d(t), t\in{\Real}^+\big).
\end{equation*}
We let $\Phi^s$ denote the locations of collaborating sensors at time $0$.

The coverage set of sensor $i$ at time $t$, denoted 
$C_i^d(\boldsymbol{\tilde{\Phi}^d}, t)$, is given by 
\begin{multline}\label{eq:rsu:temporal:coverage_set}
C_i^d(\boldsymbol{\tilde{\Phi}^d}, t)  = \\
\big\{x\in S_i^d(t) \big|x\in E_i^d(t) \mbox{~or~}l_{X_i^d(t)+Y_i, x} \cap E^{-i, d}(t) \subseteq \{x\}\big\},
\end{multline}
where $E^{-i,d}(t) = {\cup}_{j\neq i}E_j^d(t)$ is the blockage 
set associated with objects other than $i$ at time $t$.

We let $D_i^d(t)\subseteq \Realsquare$ denote sensor $i$'s region of 
interest at time $t$.
We shall define the objects that a sensor needs to sense at time $t$
as follows. 
\begin{definition} {(Objects of interest at time $t$)}
	The objects of interest of sensor $i$ at time $t$ are the 
	objects which overlap with sensor $i$'s region of interest at $t$, 
	denoted by $O_i^d(t)$, and given by
	\begin{equation}
	O_i^d(t) = \big\{j\in{\Natural}^+\big| E_j^d(t)\cap D_i^d(t)\neq \emptyset\big\}.
	\end{equation}
\end{definition}

\subsection{Sensing Redundancy and Coverage Resulting from Temporal Dynamics}
We suppose an object $i$ is sensed by object $j$ at time $t$ if 
sensor $j$ senses any part of $i$, i.e., 
\begin{equation*}
C_{j}^d(\boldsymbol{\tilde{\Phi}^d}, t)\cap E_{i}^d(t)\neq \emptyset.
\end{equation*}
Sensors can track the states of objects in the environment, e.g., 
locations, velocity, acceleration, etc, and thus have a good estimate 
of the objects even when the objects are blocked for some time. 
For simplicity we assume an object is tracked by a sensor at $t$ if 
the object has been sensed in time interval $[t-\tau, t]$, where 
$\tau$ is the maximum time window for reliable tracking without new 
sensor data. 

The spatio-temporal sensing redundancy of an object can then be defined as follows. 
\begin{definition}{(Spatio-temporal object sensing redundancy)}
	Given an environment and sensing model $\boldsymbol{\tilde{\Phi}^d}$, 
	a fixed subset of collaborating sensors, $K\subseteq \Phi^s$, and 
	assuming an object can be sensed if it has been sensed within a 
	time period $\tau$, the \emph{object sensing redundancy} of 
	sensor $i$ at time $t$ is given by
	\begin{multline}
	R^{\object,d}(\boldsymbol{\tilde{\Phi}^d}, K, i, t, \tau) = \\
	\sum_{j: X_j \in K}\indicator\big(\exists z\in [t-\tau, t] \mbox{~s.t.~} E_i^d(z)\cap C_j^d(\boldsymbol{\tilde{\Phi}^d}, z)\neq \emptyset\big).
	\end{multline}
\end{definition}

Given the above definition of spatio-temporal sensing redundancy 
we can define the $(\gamma,\tau)$-object coverage as follows. 
\begin{definition} ($(\gamma,\tau)$-object coverage)
	\label{def:rsu:temporal:object_coverage_reliability}
	Given an environment and sensing field $\boldsymbol{\tilde{\Phi}^d}$, 
	a minimum redundancy requirement $\gamma \in \Natural^+$ for reliable sensing of an object, a subset of collaborating sensors, 
	$K \subseteq \Phi^s$, and sensor $i$'s objects interest $O_i^d$, 
	the {\em $\gamma$-coverage object set} of sensor $i$ is the set of objects of interest at time $t$ which are covered by at least 
	$\gamma$ sensors in $K$, denoted by 
	\begin{equation}\label{eq:model:reliability}
	C_{c}^{d}(\tilde{\Phi}^d, K, O_i^d, \gamma, t, \tau) \delequal \big\{j\in O_i^d(t) \big| R^d(\tilde{\Phi}^d, K, j, t, \tau) \geq \gamma \big\}.
	\end{equation}
	The \emph{$(\gamma,\tau)$-object coverage} 
	is proportion of the objects of interest that are in the $\gamma$-coverage set, i.e., 
	\begin{equation}
	\frac{|C_{c}(\tilde{\Phi}, K, O_i^d, \gamma, t, \tau)|}{|O_i^d(t)|}.
	\end{equation}
\end{definition}

\subsection{Performance of Collaborative Sensing Utilizing Spatio-temporal Diversity}
The relative movement of neighboring vehicles driving in the same 
direction would typically be small, e.g., the relative locations of vehicles in a fleet may be stable most time. 
Such slow relative movement facilitates the communication amongst 
the vehicles, but limits the temporal diversity in the sensing of 
vehicles moving in the same direction.
The sensing coverage of collaborative sensing for vehicles moving 
in the same direction may fail to change quickly with time and 
obstructed vehicles will remain unseen. 
By comparison RSUs and vehicles moving in the opposite direction 
will see fast relative movements to a given flow of vehicles and 
have improved sensing coverage with temporal diversity.
We have shown in \cite{wang2018deployment} that RSUs can have an almost
unobstructed view of the road if located well above the vehicles. In practice,
RSUs may be low, e.g., to save cost, and vehicles are of different dimension,
thus the sensing of vehicles can be obstructed. However RSUs may benefit from 
temporal sensing diversity with respect to a flow of vehicles.
The relative velocity of vehicles moving in the opposite direction is large,
i.e., twice the typical speed of a vehicle, which increases temporal diversity.
However such high relative speeds can make it difficult to establish 
reliable high rate links, e.g., in the mmWave band.

Let us evaluate the performance of collaborative sensing in the presence
of such relative motions via simulation in a freeway scenario. 
We extend the simulation setting in Section~\ref{subsection:model:coverage_area_simulation}.
Sensing and communication capable RSUs are located along one side of the road
at an even spacing, denoted by $d_{\rsu}$.
The RSUs are at a distance $d_{\road}$ from the edge of the road and
the height of RSUs is $h_{\rsu}$.
Denote by $r_{\rsu}$ the sensing range of RSUs, $r_{\vehicle}$ the sensing
range of vehicles.
Both RSUs and vehicles have the same communication range $r_{\communication}$.
Vehicles are moving at the same speed $s$.
We shall refer to the direction of the lanes close to RSUs as the 
`nearby' direction, and the other direction as the `opposite' direction, 
see Fig.~\ref{fig:temporal:rsu_scenario}.

\begin{figure}
	\centering
	\includegraphics[width=1\columnwidth]{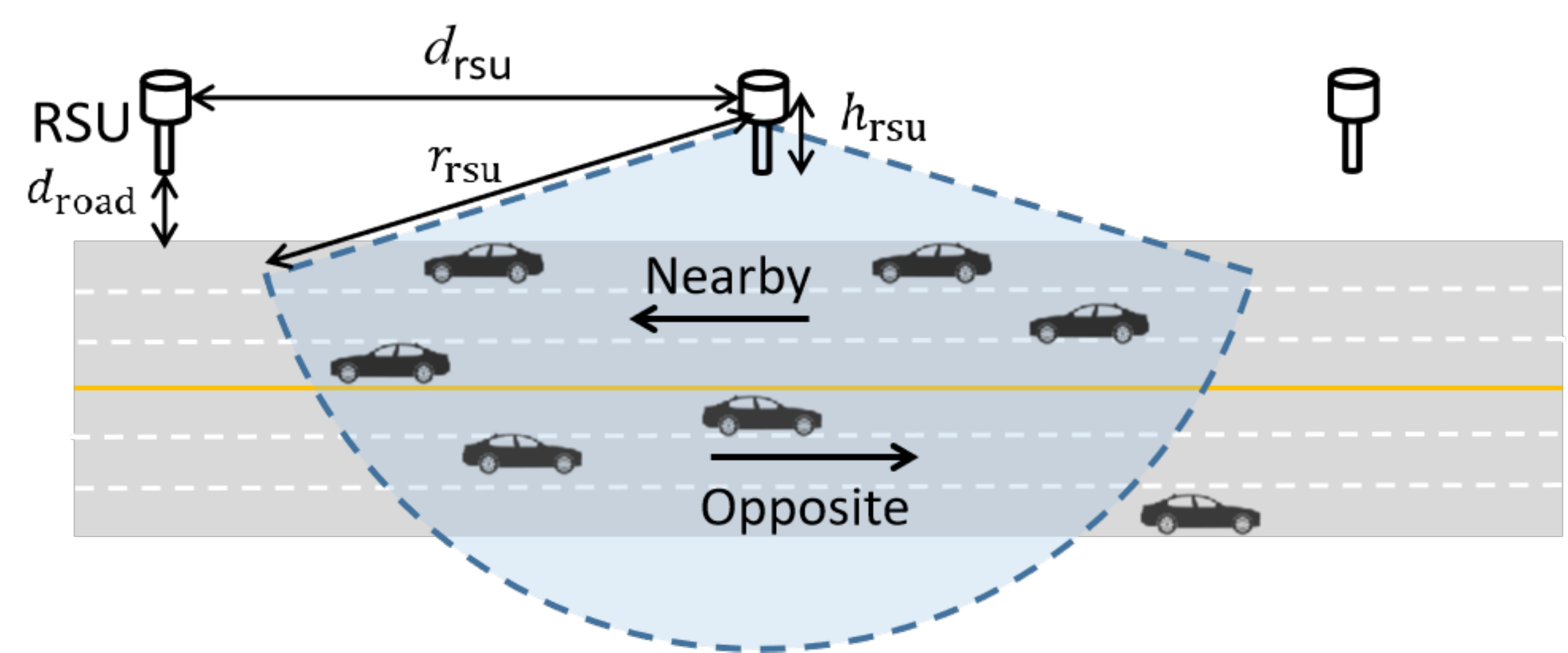}
	\caption[Freeway simulation scenario for RSU assisted collaborative sensing with temporal dynamics]
	{Freeway simulation scenario for RSU assisted collaborative sensing with temporal dynamics.}
	\label{fig:temporal:rsu_scenario}
\end{figure}

We consider different collaborative sensing schemes, i.e., 
1) base case: collaborate with only vehicles moving in the same direction. 
The communication channel is stable, yet the set of collaborating 
sensors is limited. 
2) RSU: in addition vehicles communicate with sensing capable RSUs. 
3) opposite: vehicles communicate with vehicles moving in the same 
direction and in the opposite direction. 

\begin{figure}
	\centering
	\subfloat[Collaboration with RSUs and/or vehicles in the nearby direction]{
		\includegraphics[width=0.7\columnwidth]{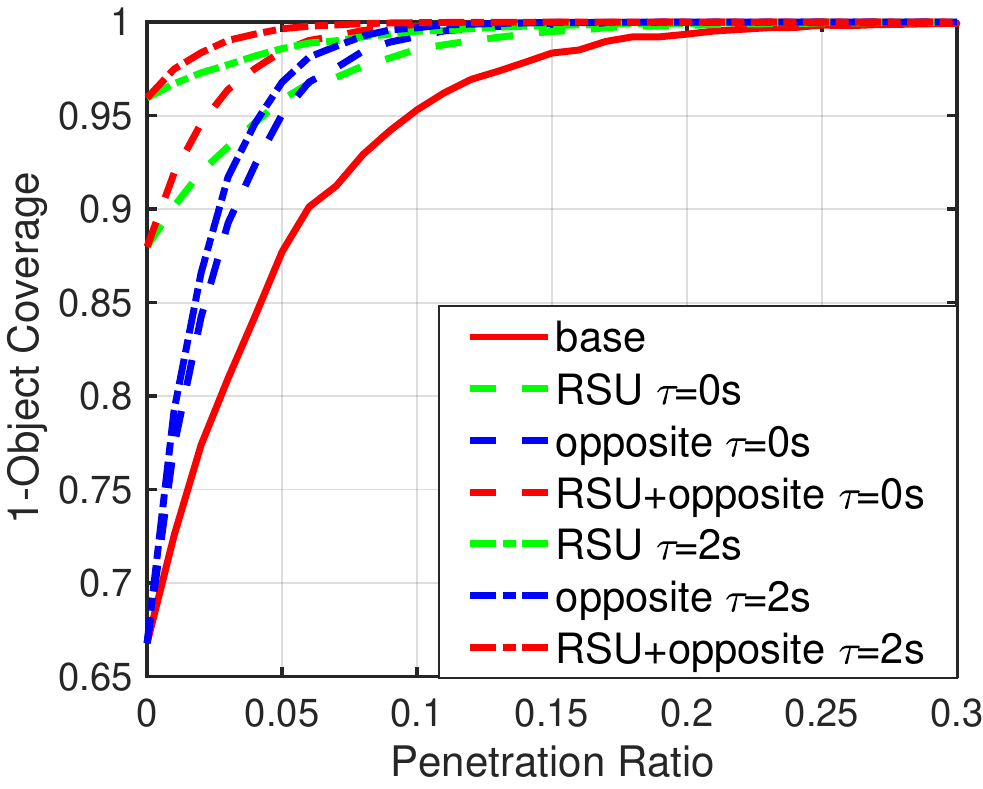}
		\label{subfig:temporal:rsu_coverage:original}
	}
	\hfill
	\subfloat[Collaboration with RSRUs and/or vehicles in the opposite direction]{
		\includegraphics[width=0.7\columnwidth]{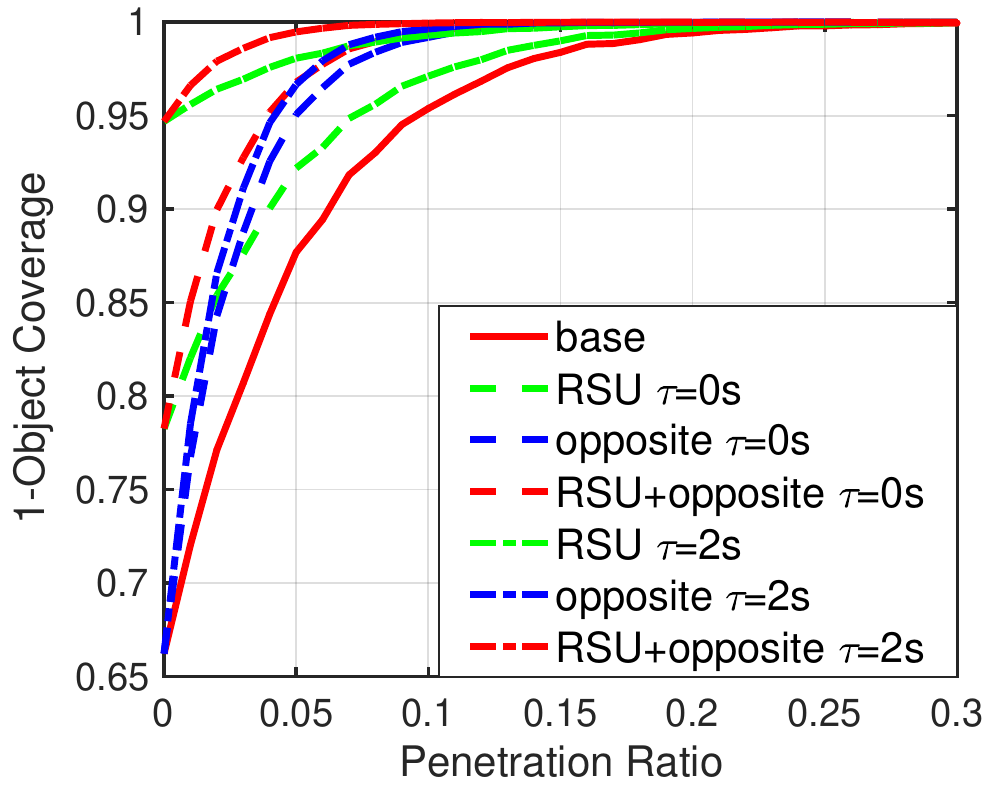}
		\label{subfig:temporal:rsu_coverage:opposite}
	}
	\caption[1-object coverage of collaborative sensing with RSUs using temporal diversity]
	{The $(1, \tau)$-object coverage of collaborative sensing with 
		vehicles driving in the same direction and RSUs for vehicles moving
		in \subref{subfig:temporal:rsu_coverage:original} the original direction, 
		and \subref{subfig:temporal:rsu_coverage:opposite} the opposite direction.
	}
	\label{fig:temporal:rsu_coverage}
\end{figure}

Fig.~\ref{fig:temporal:rsu_coverage} illustrates the $(1, \tau)$-object 
coverage of collaborative sensing under our three different
collaboration schemes and different $\tau$.
RSUs are uniformly deployed along the road, providing a $1$-coverage 
of the road, e.g., $r_{\rsu} = 200\meter$, $d_{\rsu} = 400\meter$. 
We assume $r_{\interest}= 200\meter$, $d_{\road} = 2\meter$, $r_{\vehicle} = 200\meter$, $s = 20\meter/s$.
The communication range is $r_{\communication} = 500\meter$, 
which is enough for a vehicle to communicate with all sensors having 
relevant sensor data.
We set $h_{\rsu} = 1\meter$, which is lower than the heights of vehicles, i.e.,
typically $1.5\meter$ for sedans or higher for other vehicles. 
Such assumption on $h_{\rsu}$ is mainly used to make the sensors 
subject to obstructions to study the impact of temporal diversity.
The base case is that vehicles collaborate with other vehicles 
moving in the same direction. 

First let us consider collaborative sensing without temporal diversity, i.e.,
$\tau=0\sec$. From the simulation results in Fig.~\ref{fig:temporal:rsu_coverage}
we can see sensing coverage increases with spatial diversity, 
i.e., collaboration with RSUs and/or vehicles in the opposite direction 
improves the sensing coverage. 
If we compare the coverage when only RSUs or neighbor vehicles are used, we can
see that collaborating with only RSUs provides larger gain 
at low penetrations while collaborating with neighbor vehicles 
direction works better at high penetrations.
As expected, collaborating with both RSUs and vehicles in the opposite 
direction provides most temporal diversity and thus most gain.

When temporal diversity in sensing is utilized, i.e., RSUs and 
vehicles in the opposite direction track objects using previous 
measurements, coverage can be further improved. 
In fact the coverage increases with $\tau$.
Note that collaborating with RSUs and utilizing temporal diversity 
alone can already provide a relative high coverage, e.g., over $95\%$.
This indicates that RSUs can have a good coverage of the environment by 
tracking objects even when RSUs are not located higher than all objects and are
subject to objects. A comparison of coverage for vehicles moving in different
directions shows that RSUs provide better temporal diversity for sensing
vehicles moving in the further away lanes. The reason is that the obstructions 
in the nearby lanes have larger relative movements, thus RSUs will see
larger temporal diversity in the obstruction field.

\fi

\section{Conclusion}
\label{section:conclusion}

Collaborative sensing can greatly improve a vehicle's sensing coverage.
V2V collaborative sensing could improve the sensing coverage from
$20\%$ to $80\%$ at $20\%$ penetration and we can further improve the coverage
using both V2V and V2I collaborative sensing.
However, collaborative sensing suffers at low penetrations due to, both a lack of available
collaborators, and communication blockages for (mmWave) V2V relaying paths.
Access to V2I connectivity will thus be important to provide communication
for collaborative sensing when V2V relaying paths are unavailable. 
At higher penetrations, the average V2I traffic is low, but the infrastructure 
should still have the ability to support traffic bursts when the V2V network
of collaborating vehicles becomes disconnected.

To provide higher coverage one might consider supporting joint 
collaborative sensing amongst vehicles and RSUs with both sensing 
and communication capabilities. 
With sufficient RSU density and unobstructed placements, one can ensure 
$100\%$ $1$-coverage by collaborating only with RSUs. 
The associated capacity requirement can also be much smaller than collaboration 
with vehicles: vehicles receive data from one RSU instead of all neighboring vehicles.
However sensing based only on RSUs deployed with $100\%$ $1$-coverage 
might not provide enough sensing redundancy and deploying even more 
RSUs to provide diversity would be costly. 
Furthermore, in order to navigate in a variety of environments, 
vehicles will need to have their own sensing capabilities which
should clearly be leveraged. 
Thus we see the combination of vehicular/RSU collaborative sensing 
as the most cost effective way to achieve high coverage
in vehicular automated driving applications -- in particular say for high speed
automated highways.

\appendix
\subsection{Proof of Theorem~\ref{theorem:model:coverage_area}}
The locations associated marks of the objects, $\tilde{\Phi}$,
follow an IMPPP, thus the occupied space
can be modeled by a Boolean Process \cite{chiu2013stochastic}.
One can also formally define the distribution as seen by a typical vehicle
referred to the origin $0$.
Let $Z^0 = (0, M^0)$, $M^0 = (A^0, Y^0, S^0)$, denote the typical vehicle.
We let
\begin{equation}\label{eq:proof:coverage_are:indicator_in_coverage_set}
	f(x, z^0, \tilde{\phi}\backslash \{z^0\} ) = \indicator(x\in c^0)
\end{equation}
be the indicator function that location $x$ is 
in the coverage set of the typical sensor $z^0$,
where $\tilde{\phi}\backslash \{z^0\}$ denotes the other objects in the environment excluding $z^0$.
The expected area of coverage set of a typical vehicle is then given by,
\begin{equation}\label{eq:coverage_area:reduced_palm}
	\begin{split}
		\E[|C^0|] & = {\E}\bigg[\int_{x \in \Realsquare}f(x, Z^0, \tilde{\Phi}\backslash \{Z^0\})dx \bigg]\\
			  & = \int_{x\in \Realsquare} \int_{m\in \mathbb{M}}  \int_{\tilde{\phi}}
							        f(x, \tilde{\phi}\backslash \{z\}, z) P_{z}^{!}(d\tilde{\phi})  F_{M^0}(dm)dx,
	\end{split}
\end{equation}
where $\mathbb{M}$ is the support of $M^0$, $P_{z}^{!}(\cdot)$ 
is the reduced Palm distribution of $\tilde{\Phi}$ given a typical 
object is $z=(0,m)$, i.e., the distribution of other objects 
in the environment as seen by a typical object\cite{chiu2013stochastic}.
For a Boolean Process, 
it follows by Slivnyak-Mecke theorem \cite{chiu2013stochastic} that 
the reduced Palm distribution is the same as that of the original Boolean Process.
Thus we have
\begin{equation}\label{eq:proof:coverage_area:prob_sensing_x}
\begin{split}
	& \int_{\tilde{\phi}}	f(x, z, \tilde{\phi}\backslash \{z\}) P_{z}^{!}(d\tilde{\phi}) = \int_{\tilde{\phi}}	f(x, z, \tilde{\phi}) P_{\tilde{\Phi}}(d\tilde{\phi}) \\
	\stackrel{(1)}{=} & \indicator(x\in (\{y\} \oplus s)\cap a) + \indicator(x\in (\{y\} \oplus s)\backslash a) e^{-\lambda {\E}_{A}[|l_{y , x}\oplus \check{A}|]},
\end{split}
\end{equation}
where $(\{y\} \oplus s) \backslash a$ is the sensing support of the typical sensor
excluding the region $a$ covered by the sensor itself.
In equality $(1)$ we have used the fact that for a Boolean Process, the number of objects intersecting 
a compact convex shape, e.g., $l_{y, x}$, has a Poisson distribution with 
mean $\lambda\cdot\E_{A}[|l_{y,x}\oplus\check{A}|]$, $\check{A} = \{x|-x\in A\}$ \cite{chiu2013stochastic}.
Thus substituting the result in Eq.~\ref{eq:proof:coverage_area:prob_sensing_x}
into Eq.~\ref{eq:coverage_area:reduced_palm} we get Eq.~\ref{eq:model:coverage_area:expected}.
\subsection{Proof of Theorem~\ref{theorem:collaborative:expected_rendundancy_location}}
The locations of the objects follow an HPPP and the environment can be modeled as an IMPPP
thus the environment is homogeneous in space. 
Without loss of generality we consider the redundancy of location $0$. 
By definition, we have 
\begin{equation}
\E[R(\tilde{\Phi}, \Phi^s, 0)| 0 \notin E] = \frac{\E[R(\tilde{\Phi}, \Phi^s, 0)\cdot\indicator(x\notin E)]}{\Prob(0 \notin E)}
\end{equation}
Since the region occupied by objects follows the Boolean Process 
thus the probability that $0$ is not occupied 
by objects is given by, see \cite{chiu2013stochastic},
\if \isextended 0
$\Prob(0\notin E) = e^{-\lambda\cdot \E[|A|]}$.
\else
\begin{equation}\label{eq:proof:probability_not_occupied}
\Prob(0\notin E) = e^{-\lambda\cdot \E[|A|]}.
\end{equation}
\fi
We let $h(x_0, x, m, \tilde{\phi}\backslash\{(x,m)\})$ be the indicator function 
that location $x_0$ is in the void space and sensed by object $(x,m)$, 
for the given environment excluding the reference object, i.e.,
$\tilde{\phi}\backslash\{(x,m)\}$.
$\E[R(\tilde{\Phi}, \Phi^s, 0)\cdot\indicator(0\notin E)]$ is then given by,
\begin{align}
& \E[R(\tilde{\Phi}, \Phi^s, 0)\cdot\indicator(0\notin E)] \nonumber\\
= &\E\bigg[\sum_{(X_i, M_i)\in \tilde{\Phi}, X_i \in\Phi^s } h\big(0, X_i, M_i, \tilde{\Phi}\backslash\{(X_i, M_i)\}\big)\bigg] \nonumber\\
= &p_s\lambda\int_{x\in\Realsquare}\int_{m\in\mathbb{M}}\int_{\tilde{\phi}}  h(0, x, m,\tilde{\phi})P_{(x,m)}^{!}(d\tilde{\phi})F_{M}(dm)dx\nonumber\\
\stackrel{(1)}{=} & p_s\lambda\int_{x\in\Realsquare} {\E}_{M,\tilde{\Phi}}\big[h(0, x, M,\tilde{\Phi})\big]dx \nonumber\\
\stackrel{(2)}{=} & p_s\lambda\int_{x\in\Realsquare} {\E}_{M,\tilde{\Phi}}\big[h(-x, 0, M,\tilde{\Phi})\big]dx \nonumber\\
\stackrel{(3)}{=} &  p_s\lambda\cdot \E[|C^0\backslash A^0|].
\label{eq:proof:expected_redundancy_of_void}
\end{align}
The equality $(1)$ follows for Slivnyak-Mecke theorem \cite{chiu2013stochastic}.  
Equality $(2)$ follows from the spatial homogeneity of the environmental model
thus we have that \if\isextended0
${\E}_{M,\tilde{\Phi}}\big[h(0, x, M,\tilde{\Phi})\big] = {\E}_{M,\tilde{\Phi}}\big[h(-x, 0, M,\tilde{\Phi})\big].$
\else$${\E}_{M,\tilde{\Phi}}\big[h(0, x, M,\tilde{\Phi})\big] = {\E}_{M,\tilde{\Phi}}\big[h(-x, 0, M,\tilde{\Phi})\big].$$\fi
Equality $(3)$ follows from the result characterizing $\E[|C^0|]$ in Thm.~\ref{theorem:model:coverage_area}. 
Note that function $h$ is not the same as $f$ introduced in the proof of Thm.~\ref{theorem:model:coverage_area}, 
i.e., a point on the boundary of an object 
can be in the coverage set but can not be in the void space. 
However the area of the set of such points is $0$, thus equality $(3)$ holds.
Combining the above results finishes the proof.
\if \isextended 1
\subsection{Proof of Theorem~\ref{theorem:scalability:v2i:1d_capacity}}
We shall consider the expected number of V2I transmissions required by a
typical sensing vehicle. 

\emph{V2I uplink.}
The probability that the V2I link will be required to share sensor data with
collaborating vehicles in one direction, e.g., forward direction, is given by
\begin{equation}
        {p}_{\front}(\eta, p_s) = 1 - \sum_{k = 0}^{\eta} p_s^{k}\cdot(1-p_s)^{\eta - k}.
\end{equation}
This expression can be interpreted as one minus the probability (associated
with the sum) that the V2I link is not required. The V2I link will not be required
if the first $k$ vehicles are collaborative and can thus perform V2V relaying,
and the remaining $\nu-k$ are not sensing vehicles and so do not require the data.
The forward and backward directions are independent
and symmetric, thus  the probability that V2I resources will be required is
\begin{equation}\label{eq:scalability:v2i:1d:p_v2i}
        {p}_{\vtoi}(\eta, p_s) = 1 - \big(1 - {p}_{\front}(\eta, p_s)\big)^2.
\end{equation}
Note that data need only be sent up once irrespective of whether
one or more sharing paths are blocked thus
$\E[N_{\uplink}] = p_{\vtoi}$.

\emph{V2I downlink.}
If broadcast downlink is used, we have $N_{\downlink}^{\broadcast} = N_{\uplink}$,
thus $\E[N_{\downlink}^{\broadcast}] = \E[N_{\uplink}]$.
If only a unicast downlink is available, a V2I downlink is
required for every collaborative vehicle where no LOS V2V relay path is available.
Given our modeling assumption that vehicles receiving data from infrastructure can
further relay data via V2V links, the $(k+1)^{th}$ collaborative
vehicle requires a downlink transmission if the $k^{th}$ vehicle is not sensing.
$\E[N_{\downlink}^{\unicast}]$ is thus the sum of the expected number of 
unicast downlink transimission required by each $k^{th}$ vehicle and thus we
have Eq.~\ref{eq:scalability:v2i:1d:n_dl_u}.

Given the expected numbers of V2I transmissions for a typical sensing vehicle, 
we get the associated capacity $c_{\uplink}$, $c_{\downlink}^{\broadcast}$, and $c_{\downlink}^{\unicast}$ accordingly.
\fi
\if \isextended 1	
	\subsection{V2I Capacity with Assistance from Neighbor Lanes}

Consider the vehicles in front of a reference vehicle placed in
column $0$ of the grid. 
Let $S_k = (S_k^1, S_k^2, S_k^3)$, $S_k^i\in\{0,1\}$, denote whether 
the vehicles in the $k^{th}$ column from the reference vehicle
($1,2,3$ denotes vehicles from top row to bottom row) 
are collaborating where
$0$ denotes a non-collaborating vehicle and $1$ the opposite. 
Denote by $X_k = (X_k^1, X_k^2, X_k^3)$, $X_k^i\in \{0,1\}$, 
the state of the vehicles in the $k^{th}$ column are both 
collaborating and can receive data from the reference vehicle.
We denote by $Y_k \in \{0, 1\}$ whether the V2I downlink is 
required to relay sensing data to vehicles in the \emph{first} $k$ columns. 
The state of the $k^{th}$ column is given by
\begin{equation}
Z_k = (X_k, Y_k).
\end{equation}

Based on our assumption that relaying paths can
not contain links in both forward and backward directions,
$X_{k+1}$ only depends on $X_{k}$ and $S_{k+1}$.
Since whether a vehicle is collaborating is independent 
from other vehicles, the probability distribution of $Z_{k+1}$
depends on that of $Z_{k}$ and $p_s$.

Denote by $P$ the state transition probability of a transition from $Z_k$ to
$Z_{k+1}$, $k\geq 0$. The probability distribution of $S_{k+1}$ is given by, 
\begin{equation}
\Prob\big(S_{k+1} = (s_k^1, s_k^2, s_k^3)\big) = p_s^{s_k^1 + s_k^2 + s_k^3}\cdot (1-p_s)^{3 - s_k^1 - s_k^2 - s_k^3}.
\end{equation}
Denote by $\tilde{X}_{k+1}$ the indicator that vehicles in 
the $k^{th}$ column can send data to vehicles in the $(k+1)^{th}$ column via V2V links. 
Denote by $\wedge$ a logical AND, and by $\vee$ a logical OR, then we have that
\begin{equation}
\tilde{X}_{k+1} = (X_k^1 \wedge S_{k+1}^1, X_k^2 \wedge S_{k+1}^2, X_k^3 \wedge S_{k+1}^3).
\end{equation}
Further consider the communication amongst vehicles in the same column.
Denote by $\hat{X}_{k+1}$ the state of vehicles after vehicles in the 
$(k+1)^{th}$ column share data amongst themselves via V2V links; then we have
\begin{align}
\hat{X}_{k+1}^1 & = \tilde{X}_{k+1}^1 \vee ( S_{k+1}^1 \wedge (\tilde{X}_{k+1}^2 \vee (\tilde{X}_{k+1}^3 \wedge S_{k+1}^2))), \\
\hat{X}_{k+1}^2 & = \tilde{X}_{k+1}^2 \vee (S_{k+1}^2 \wedge (\tilde{X}_{k+1}^1 \vee \tilde{X}_{k+1}^3) ), \\
\hat{X}_{k+1}^3 & = \tilde{X}_{k+1}^3 \vee ( S_{k+1}^3 \wedge (\tilde{X}_{k+1}^2 \vee (\tilde{X}_{k+1}^1 \wedge S_{k+1}^2))),
\end{align}
i.e., a sensing vehicle can also receive data from 
other collaborating vehicles in the same column via V2V relaying.

For V2I relaying, we denote by $\tilde{Y}_{k+1}$ whether 
V2I relaying is required by the $k+1$th column. 
This occurs if the vehicle in the central lane is collaborating but
can not receive data via V2V links, i.e., when
\begin{equation}\label{eq:scalability:v2i:2d:v2i_condition}
(S_{k+1}^2 = 1) \mbox{~and~} (\hat{X}_{k+1}^2 = 0),
\end{equation}
we have $\tilde{Y}_{k+1} = 1$. 
The vehicle can further relay data to neighboring collaborative vehicles 
in the $(k+1)^{th}$ column. The state transition is now given by
\begin{equation}
Y_{k+1} = Y_{k} \vee \tilde{Y}_{k+1}, 
\end{equation}
\begin{equation}\label{eq:scalability:v2i:2d:final_transition}
X_{k+1} =  
\begin{cases}
\hat{X}_{k+1} ,& \text{if } \tilde{Y}_{k+1} = 0\\
S_{k+1},              & \text{otherwise}
\end{cases}.
\end{equation}

Based on the above state transition rules, we can compute $P$ as a function of $p_s$. 
Denote by $\mathcal{Z}$ the support of $Z_k$, 
$\pi_k = (\pi_k^1, \pi_k^2, \ldots, \pi_k^{|\mathcal{Z}|})$ the probability distribution of $Z_k$, where $\pi_k^i$ is the probability of state $i$ at column $k$. 
We have that
\begin{equation}
\pi_k = P^k \cdot \pi_0.
\end{equation}
Denote by $\mathcal{Z}_{\vtoi}\subseteq \mathcal{Z}$ the set of states with $Y=1$. 
The probability that V2I communication is required to relay data to 
vehicles in the forward direction, conditioning on the probability distribution 
of column $Z_0$ being $\pi_0$, is given by 
\begin{equation}
	p_{\front}(\eta, p_s, \pi_0) = \sum_{i \in \mathbb{Z}_{\vtoi}} \pi_{\eta}^i,
\end{equation}
where $\pi_{\eta} = P^{\eta}\pi_0$.
Conditioning that the reference vehicle is a sensing vehicle, 
we can compute $\pi_0$ based on $p_s$. 
$p_{\vtoi}$ is thus given by 
\begin{equation}
p_{\vtoi} = \sum_{i = 1, \ldots, |\mathcal{Z}|} \pi_0^i \cdot \big(1 - (1 - p_{\front}(\eta, p_s, e_i))^2\big),
\end{equation}
where $e_i\in \{0,1\}^{|\mathcal{Z}|}$, $e_i^i = 1$ and $e_i^j = 0$ for $j\neq i$.

For the number of $\E[N_{\downlink}^{\unicast}]$, we can define $Y_k$ as the state for number of V2I unicast downlinks required by vehicles in each column.
Similarly as above, we can compute the corresponding state transition probability and  $\E[N_{\downlink}^{\unicast}]$ is given by
\begin{equation}
	\E[N_{\downlink}^{\unicast}] = \sum_{k = 1}^{\eta}\E[Y_k]
\end{equation}

In the above analysis we assume the reference vehicle only needs to 
share data to vehicles in the same lane, e.g., vehicles are
moving in platoons and mainly require data from the same platoon. 
In fact, vehicles may also need to share data with vehicles in neighboring
lanes for applications such as advanced automated driving and collaborative sensing \cite{3gpp2017v2xstage1}. 
In this case we can analyze the required capacity
on V2I network following similar steps. One major difference is that 
the condition in Eq.~\ref{eq:scalability:v2i:2d:v2i_condition} should be replaced by
\begin{equation}
\exists i\in \{1,2,3\} \mbox{~s.t.~} \hat{X}_{k+1}^i \neq S_{k+1}^i,
\end{equation}
i.e., there is a sensing vehicle not receiving the sensor data via V2V relay.
Also in Eq.~\ref{eq:scalability:v2i:2d:final_transition} we have $X_{k+1} = S_{k+1}$, 
i.e., all sensing vehicles would get the data by either V2V or V2I.

\begin{figure}
	\centering
	\includegraphics[width=\figurewidth\columnwidth]{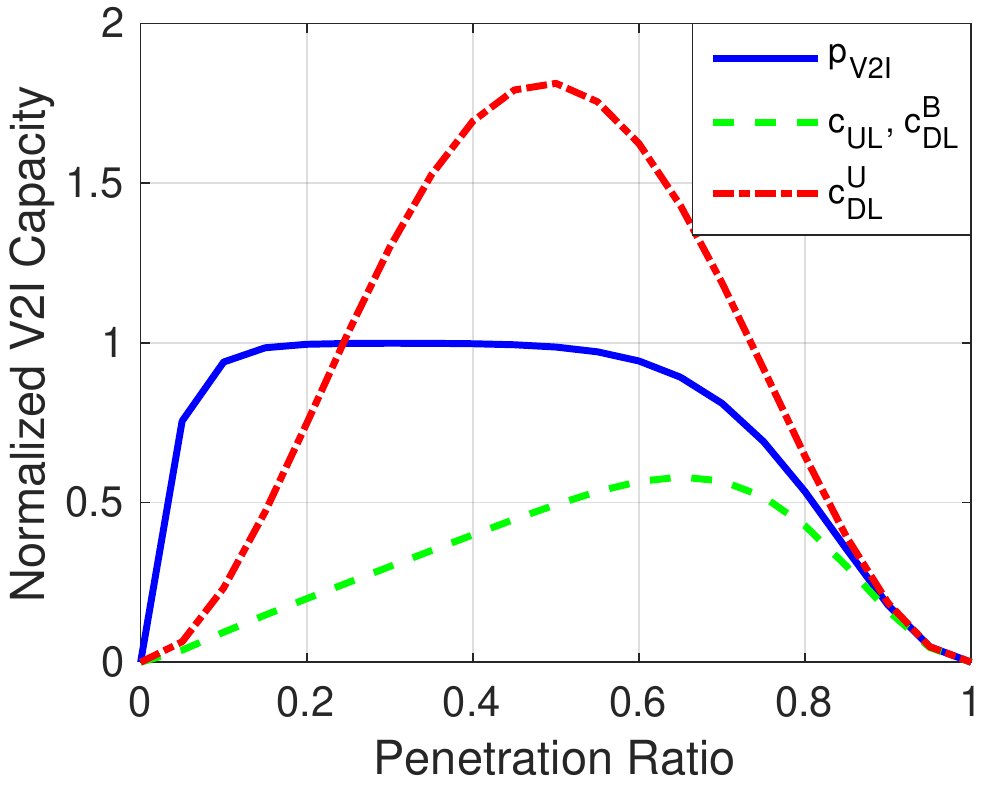}
	\caption[Scalability of V2I capacity in 2-D grid model]{How $p_{\vtoi}(\eta,p_s)$, 
		normalized $c_{\uplink}$, $c_{\downlink}^{\broadcast}$, and $c_{\downlink}^{\unicast}$
		change with $p_s$ when vehicles send data to vehicles in the same lane 
		and the two neighboring lanes. }
	\label{fig:scalability:v2i:2d:capacity_all_vehicles}
	\vspace{-0.2in}
\end{figure}

In Fig.~\ref{fig:scalability:v2i:2d:capacity_all_vehicles} we exhibit the 
result when vehicles need to share data with vehicles on neighboring lanes.
Compared with the case that vehicles need to share data with only vehicles in the 
same lane, the V2I capacity requirements here is much higher. 
Such a result was to be expected as more vehicles require sensing data. 
Note that $p_{\vtoi}$ is almost $1$ for a large range of penetrations, e.g., 
from $0.1$ to $0.6$.
This indicates that assistance from V2I would be necessary for reliable 
collaborative sensing from the early stages when the penetration of automated
driving vehicles is low.
\fi

\bibliographystyle{IEEEtran}  
\bibliography{yw}        
\index{Bibliography@\emph{Bibliography}}%

\end{document}